\documentclass[10pt,twocolumn,letterpaper]{article}

\usepackage[pagenumbers]{cvpr} 

\usepackage{xspace}
\usepackage{multirow}
\usepackage{multicol}

\definecolor{caribbeangreen}{rgb}{0.0, 0.8, 0.6}
\definecolor{coralred}{rgb}{1.0, 0.25, 0.25}

\definecolor{royalblue(web)}{rgb}{0.25, 0.41, 0.88}
\definecolor{jade}{rgb}{0.0, 0.66, 0.42}
\definecolor{jasper}{rgb}{0.84, 0.23, 0.24}
\definecolor{amber}{rgb}{1.0, 0.75, 0.0}

\newcommand{\gen}{\mathcal{G}\xspace} 
\newcommand{\proj}{P\xspace} 
\newcommand{\inversion}{\mathcal{E}\xspace} 
\newcommand{\dimension}{d\xspace} 
\newcommand{\optenc}{\textbf{O}\xspace} 
\newcommand{\digenc}{\mathcal{D}_{\theta}\xspace} 
\newcommand{\latentplus}{\textbf{L}\xspace} 
\newcommand{\img}{\textit{I}\xspace} 
\newcommand{\imgo}{\textit{I'}\xspace} 
\newcommand{\imgc}{\textit{J}\xspace} 
\newcommand{\gendimension}{k\xspace} 
\newcommand{\gendimensiond}{l\xspace} 

\definecolor{cvprblue}{rgb}{0.21,0.49,0.74}
\usepackage[pagebackref,breaklinks,colorlinks,allcolors=cvprblue]{hyperref}

\usepackage{booktabs}
\usepackage{array}
\usepackage{graphicx}
\title{Latent Space Imaging}

\author{
    Matheus Souza \hspace{2em} Yidan Zheng \hspace{2em} Kaizhang Kang \hspace{2em} Yogeshwar Nath Mishra \\
    \vspace{1em}
    \hspace{6em} Qiang Fu \hspace{4em} Wolfgang Heidrich \\
    KAUST \\
}

\begin{document}

\makeatletter
\let\@oldmaketitle\@maketitle
\renewcommand{\@maketitle}{\@oldmaketitle
    \includegraphics[width=1.0\textwidth]{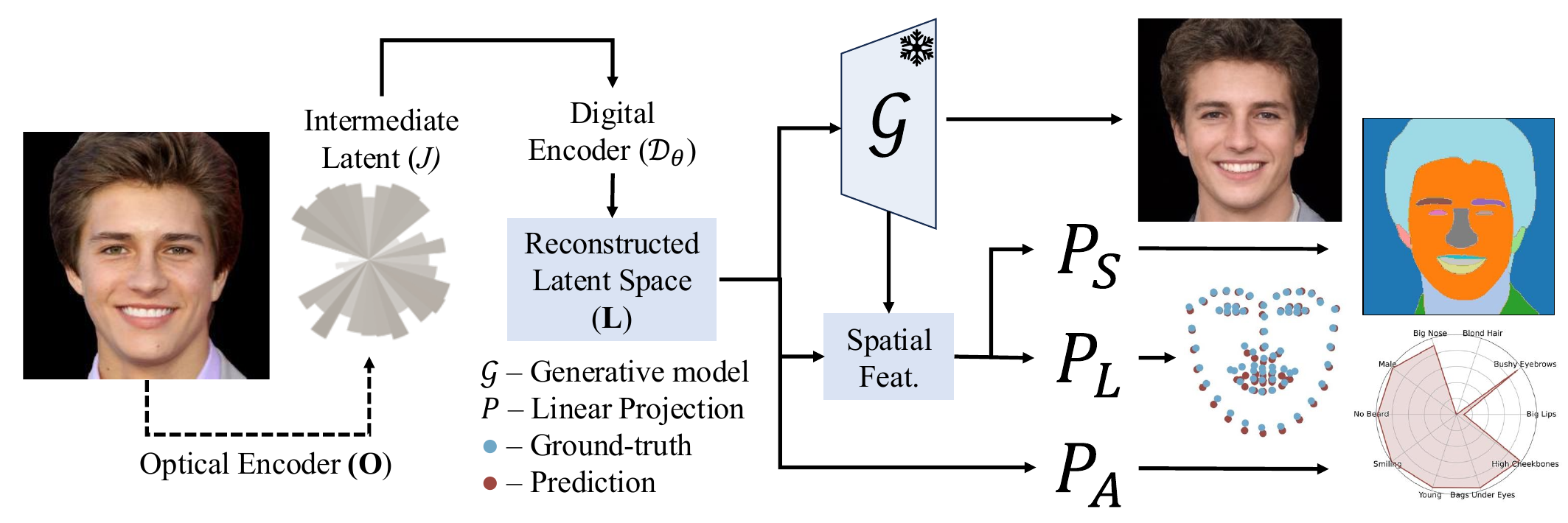}
    \centering
     \captionof{figure}{We propose an extremely-compressed imaging paradigm called Latent Space Imaging (LSI). The optical encoder (\optenc) projects the real signal into a compressed set of measurements. A digital encoder ($\digenc$) then maps this signal to the latent space (\latentplus) of a frozen generative model ($\gen$), enabling image reconstruction. The \latentplus can also be linearly projected ($\proj$) to perform downstream tasks directly—such as facial segmentation ($\proj_{S}$), landmark detection ($\proj_{L}$), and attribute classification ($\proj_{A}$)—without requiring image reconstruction or a complex new model.}
    \label{fig:teaser}
\bigskip}

\makeatother

\maketitle

\begin{abstract}
Digital imaging systems have traditionally relied on brute-force measurement and processing of pixels arranged on regular grids. In contrast, the human visual system performs significant data reduction from the large number of photoreceptors to the optic nerve, effectively encoding visual information into a low-bandwidth latent space representation optimized for brain processing. Inspired by this, we propose a similar approach to advance artificial vision systems.

Latent Space Imaging introduces a new paradigm that combines optics and software to encode image information directly into the semantically rich latent space of a generative model. This approach substantially reduces bandwidth and memory demands during image capture and enables a range of downstream tasks focused on the latent space.

We validate this principle through an initial hardware prototype based on a single-pixel camera. By implementing an amplitude modulation scheme that encodes into the generative model's latent space, we achieve compression ratios ranging from 1:100 to 1:1000 during imaging, and up to 1:16384 for downstream applications. This approach leverages the model's intrinsic linear boundaries, demonstrating the potential of latent space imaging for highly efficient imaging hardware, adaptable future applications in high-speed imaging, and task-specific cameras with significantly reduced hardware complexity.
\end{abstract}
    
\section{Introduction}\label{sec1}

Conventional imaging systems are based on brute-force sampling of
scene information on regular grids. In contrast, biological vision systems operate
very differently: for example the human visual system has around 120
million rods and 7 million cones~\cite{molday2015photoreceptors}, but the information from these
photoreceptors is distilled by retinal processing into neural signals
that are transmitted over the optic nerve with only 0.7-1.7 million
axions~\cite{jonas1992human}. Therefore, our vision employs compression to encode image information into a latent
space compatible with the neural architecture of the human
brain.

Here we propose Latent Space Imaging (LSI), a new paradigm that aims
to bring this concept to the realm of cameras and artificial vision
systems. Specifically, we propose to utilize the combination of linear
optical encoding for drastic data reduction, paired with minimal non-linear computational processing to directly encode image
information into the latent space of a generative model. This approach extremely reduces the amount of sensor data required for downstream tasks such as face segmentation,
facial landmark detection, or image reconstruction.

To showcase this framework, we focus here on the rich
latent space of a generative model in the form of a
StyleGANXL~\cite{karras2019stylebased,Karras2019stylegan2,karras2021alias,sauer2022stylegan}
trained on human faces. StyleGAN models are renowned for their ability to
synthesize highly realistic face images, while supporting editing and
image-to-image translation applications. Their semantically rich, disentangled, well-structured, and \emph{compact} latent space makes them an ideal architecture for LSI. This structured representation facilitates key downstream applications, including landmark detection, facial feature segmentation, and the extraction of facial attributes such as age, gender, smile/no smile, and beard/no beard. Furthermore, we demonstrate its effectiveness in the particularly challenging task of reconstructing full human face images from a very small number of samples.

While there are a number of possible ways to optically implement the
LSI paradigm, our first prototype utilizes the single-pixel imaging
(SPI)
concept~\cite{duarte2008single,katz2009compressive,song2023singlepixel,zhang2015single}. A
discussion of alternative implementations~\cite{wetzstein2020inference,chang2018hybrid,lin2018all,wei2023spatially} can be found in the Supplement.

In summary, we make the following contributions:
\begin{itemize}
  \item the introduction of Latent Space Imaging as a new paradigm for
    very low bandwidth image capture;
  \item a proof-of-concept hardware prototype for LSI utilizing the SPI
    framework, featuring a programmable optical system for easy and
    extensive experimentation,
  \item an initial latent space for human faces based on a StyleGAN's family of
    generative model; and

  \item the demonstration of a range of downstream applications on this latent space with real hardware experimentation.
\end{itemize}

We also refer the readers to the Supplement for a detailed discussion of hardware and software choices for the LSI paradigm. Our implementation is available at \url{https://github.com/vccimaging/latent-imaging}.

\section{Related Work}\label{sec2:related}
\subsection{Optics and Algorithm Co-design}

Latent Space Imaging builds and improves upon the recent trend for
co-designing both optics and reconstruction algorithms. This joint
optimization has yielded favorable outcomes in computational imaging
and downstream computer vision tasks. These include color image
restoration~\cite{Chakrabarti}, microscopy~\cite{NehmeDeepSTORM3D,
  KellmanPtychographic, ShechtmanMulticolour}, monocular depth
imaging~\cite{chang2019deep, haim2018depth, he2018learning,
  wu2019phasecam3d}, super-resolution, extended depth of
field~\cite{sitzmann2018end, SunRayTracing2021}, time-of-flight
imaging~\cite{chugunov2021mask}, and high-dynamic range
imaging~\cite{sun2020learning}, among others. The enhancements
observed are attributed to the ability of such systems to
\emph{optically encode} relevant scene information into sensor data,
with the computational post-processing acting as a \emph{decoder}. The
crucial innovation in LSI is that the code represents the semantically
expressive yet efficiently encoded latent space of a generative model,
with that model acting as the decoder module.

\subsection{Compressive Sensing and Minimal Cameras}

Our approach aligns with the principles of Compressive Sensing (CS)~\cite{donoho2006compressed}, which harnesses sparsity and structural priors to recover signals from minimal samples. Traditional CS methods emphasize optimization-based reconstruction, while recent advancements integrate deep learning, combining sparsity-driven priors with neural network capabilities. This first group of methods has led to diverse architectures, including CNNs, transformers, and deep unfolding models~\cite{zhang2018ista,mou2022deep,song2023optimization,wang2023saunet,Qu_2024_CVPR,wang2024ufc,wang2024hierarchical}, bridging the gap between classical CS and data-driven approaches. Further, generative priors, forming a second group, offer a robust alternative by leveraging latent spaces to enhance image quality and structure~\cite{bora2017compressed,kabkab2018task,jalal2020robust,kelkar2021compressible,kelkar2021prior}, showing the potential of learned priors to capture dependencies within compressed data. A third group advances these concepts in real-world applications through single-pixel imaging~\cite{duarte2008single,katz2009compressive,song2023singlepixel,zhang2015single,higham2018deep,Huang23,s19194190,yang2021high}, using quantized patterns, binary random masks, Fourier domain methods, and co-optimized mask-reconstruction pipelines. Lastly, a fourth group~\cite{klotz2025minimalist,pooj2018minimalist} presents minimalist cameras with optimized freeform pixels for task-specific applications like workspace monitoring.

\textit{Latent Space Imaging} stands apart from both the first and
second groups by emphasizing realistic physical deployment, requiring
attention to numerical constraints, quantization, light efficiency,
and similar practical elements. Unlike the second group, which often
employs iterative optimization, our approach reconstructs images in a
single forward pass. Additionally, we distinguish our work from these
groups by targeting latent space reconstruction itself. Our framework
uniquely leverages once-optimized masks for multi-model purposes,
avoiding the conventional use of latent space solely as an
optimization prior and without optimizing the masks/pixels for
multiple tasks. Finally, we set ourselves apart from the fourth
group design by utilizing a latent space
to enable fine-grained tasks, such as face identification, with a
smaller number of \textit{freeform} samples than traditional cameras.

\section{Latent Space Imaging}\label{sec3}

Traditional imaging captures a regular $m \times n$ grid of measurements, while compression strategies typically use a collection of random linear codes to leverage the information redundancy inherent in real images in combination with a reconstruction software
module. In contrast, we demonstrate that meaningful information can be significantly more compressed in the latent space of generative models while providing sufficient information for downstream processing tasks.

As depicted by Fig.~\ref{fig:teaser}, our objective is to learn an optically encoded mapping 
\begin{equation}\label{eq:mapping}
\begin{matrix}
\latentplus = \digenc(\optenc \cdot \img),
\end{matrix}
\end{equation}
\noindent where the input image $\img \in \mathbb{R}^{mn}$ is mapped
to a tensor $\latentplus \in \mathbb{R}^{\gendimension \times l}$ representing
the latent space of a generative model. This mapping is achieved
through the combination of a co-designed (linear) optical encoder
$\optenc \in \mathbb{R}^{\dimension \times mn}$ and a digital
(nonlinear) encoder $\digenc$, with $\theta$ representing the set of
parameters that configure the digital encoder. Where $\dimension$ is the dimension of the compressed measurement, $l$ is the number levels from coarse to fine inside the generative models as well as in the latent space, and $k$ represents latent dimension. 

The optical encoder linearly projects the image to an intermediate
vector $\imgc \in \mathbb{R}^\dimension$ ($\optenc \cdot \img$) representing the actual
measurements of the signal. This substantial compression ratio is achieved by optimizing
the projection alongside other components during the image
reconstruction process. Instead of merely matching pixel values, the
process targets the generative model's latent representation
$\latentplus$ of the original signal. By focusing on this compressed
space, the essential features of the data distribution within a
particular domain are captured, effectively eliminating redundancies
inherent in the data.

However, due to the complex non-linear operations intrinsic to
generative models and the impracticality of realizing these
non-linearities optically, an additional digital encoder $\digenc$ is
necessary to align $\imgc$ with the entire latent space of
interest. This alignment is achieved through a deep neural network
(DNN), that enhances the representation by effectively addressing the
non-linearities. The digital encoder $\digenc$ expands $\imgc$ to
$\latentplus \in \mathbb{R}^{512 \times 18}$ (see
Fig.~\ref{fig:teaser}), aligning it with the dimensions of the chosen
generative model. In our work, StyleGANXL~\cite{sauer2022stylegan} is employed since it performs robustly within specific domains, and its mechanisms for inversion to latent space are well-studied~\cite{xia2022gan}.

Specifically, the digital encoder employs a multi-level network inspired by advancements in the inversion
domain~\cite{richardson2021encoding}. This approach leverages the detailed structural levels of StyleGANXL. In
this framework, the optically encoded vector $\imgc$ passes through a series of linear layers and attention mechanisms, each followed by activation functions. The depth of these stacks varies to match the required level of detail. For the pre-trained StyleGANXL on human
faces, there are 18 levels corresponding to different resolutions. The
number of stacks increases progressively from coarse to fine, aligning
with low to high resolutions, respectively.

After acquiring all detail levels, we obtain $\latentplus \in \mathbb{R}^{512 \times 18}$. A mixer block, inspired by \cite{liu2021pay}, is then used to learn a weighted mixture of information across these levels, which is subsequently fed into the generative model $\gen$ to output the image $\imgo$

\begin{equation}\label{eq:gen}
\begin{matrix}
\imgo = \gen(\latentplus),
\end{matrix}
\end{equation}

\noindent where $\imgo \in \mathbb{R}^{mn \times 3}$, with $m$ and $n$
determined by the capabilities of the generative model $\gen$. The
value 3 corresponds to the RGB color channels which are discussed in
further detail in the Supplement. 

A pivotal aspect of the LSI framework is the identification of the optimal latent space to serve as ground-truth during the optimization process. To accomplish this, we trained an encoder, $\inversion$, following the method proposed by \cite{richardson2021encoding} for accessing the StyleGANXL latent space. This encoder provides an approximate ground-truth latent representation, $\inversion(\img)$, which serves as the target for optimizing each entry of the $\optenc$ matrix and the $\digenc$. The optimization is driven by the loss function $\mathcal{L}_{lat}$

\begin{equation}
\begin{matrix}
\mathcal{L}_{lat} = \left \| \digenc(\optenc \cdot \img)) - \inversion(\img) \right \|_1.
\end{matrix}
\label{eq:minimize}
\end{equation}

Making the encoding $\latentplus$ close to $\inversion(\img)$ guarantees that it is inside the generative model average space and diminishes
hallucination caused by only matching an $\ell_2$ norm at the pixel level.
To further enhance facial fidelity, we incorporate additional terms in the loss function. Detailed information on these loss functions can be found in the Supplement.

\section{Linear Learning for Downstream Tasks}
\label{sec:semantics}

In the LSI pipeline, we use a hybrid encoding setting to generate $\latentplus$, enabling high-level vision tasks with minimal training and a very limited number of measurements. Previous research has examined the semantics of GAN models, uncovering latent space linear boundaries that distinguish different image attributes~\cite{xu2021linear,collins2020editing,shen2020interpreting,Abdal_2019_ICCV}. In particular, \cite{xu2021linear} demonstrated that linear transformations can efficiently extract semantic information from feature maps without the need for image reconstruction or specialized encoders for various tasks.

Our approach involves performing three different high-level facial
analysis tasks using simple linear transformations {\em from a single
  latent space} $\latentplus$. Unlike prior studies, we push
compression limits by assessing the minimum measurements needed to
derive meaningful facial semantics through linear projections $\proj$ from the
same latent space. Once $\optenc$ and $\digenc$ are optimized,
they are kept frozen and different $\proj$ are optimized for downstream applications such as:

\noindent\textbf{Attribute Classification} predicts facial attributes (\textit{e.g.}, age, beard, smile, gender) using a linear feature mapper $\proj_{A}$, which is implemented as a fully connected layer that projects $\latentplus \in \mathbb{R}^{512 \times 18}$ to $\mathbb{R}^{40}$, representing 40 distinct facial characteristics. \\
\textbf{Face Segmentation} provides pixel-wise segmentation across facial regions (\eg, nose, eyes, face, mouth). To accurately capture spatial structure, we employ multi-level feature extraction from the generative model, combining these features through a linear projection block, $\proj_{S}$, which utilizes bilinear interpolation alongside convolutional layers. The generative model $\gen$ offers feature maps at various resolutions, spanning from coarse ($36 \times 36$) to fine ($256 \times 256$). We extract features across six distinct levels, representing a comprehensive range of resolutions. This multi-level feature integration allows for precise semantic pixel assignment, effectively addressing the spatial limitations of raw \latentplus features, which alone lack the fine-grained detail essential for high-accuracy segmentation.
\\
\textbf{Landmark Detection} yields the coordinates of key facial points within an image. Unlike segmentation, where multiple features are extracted, we focus here on a single coarse feature map from the generative model to capture spatial structures. This coarse feature is denoted by $\in \mathbb{R}^{1024 \times 36 \times 36}$ and is modulated by the latent space, \latentplus itself. A convolutional layer, combined with bilinear interpolation, projects and downsamples this feature to $\mathbb{R}^{68 \times 8 \times 8}$. Finally, a linear layer reduces the spatial dimensions, representing the 68 key points as 2D coordinates.

\section{Hardware Prototype} \label{sec:resphy}
\begin{figure}[!ht]
    \centering
    \includegraphics[width=0.45\textwidth]{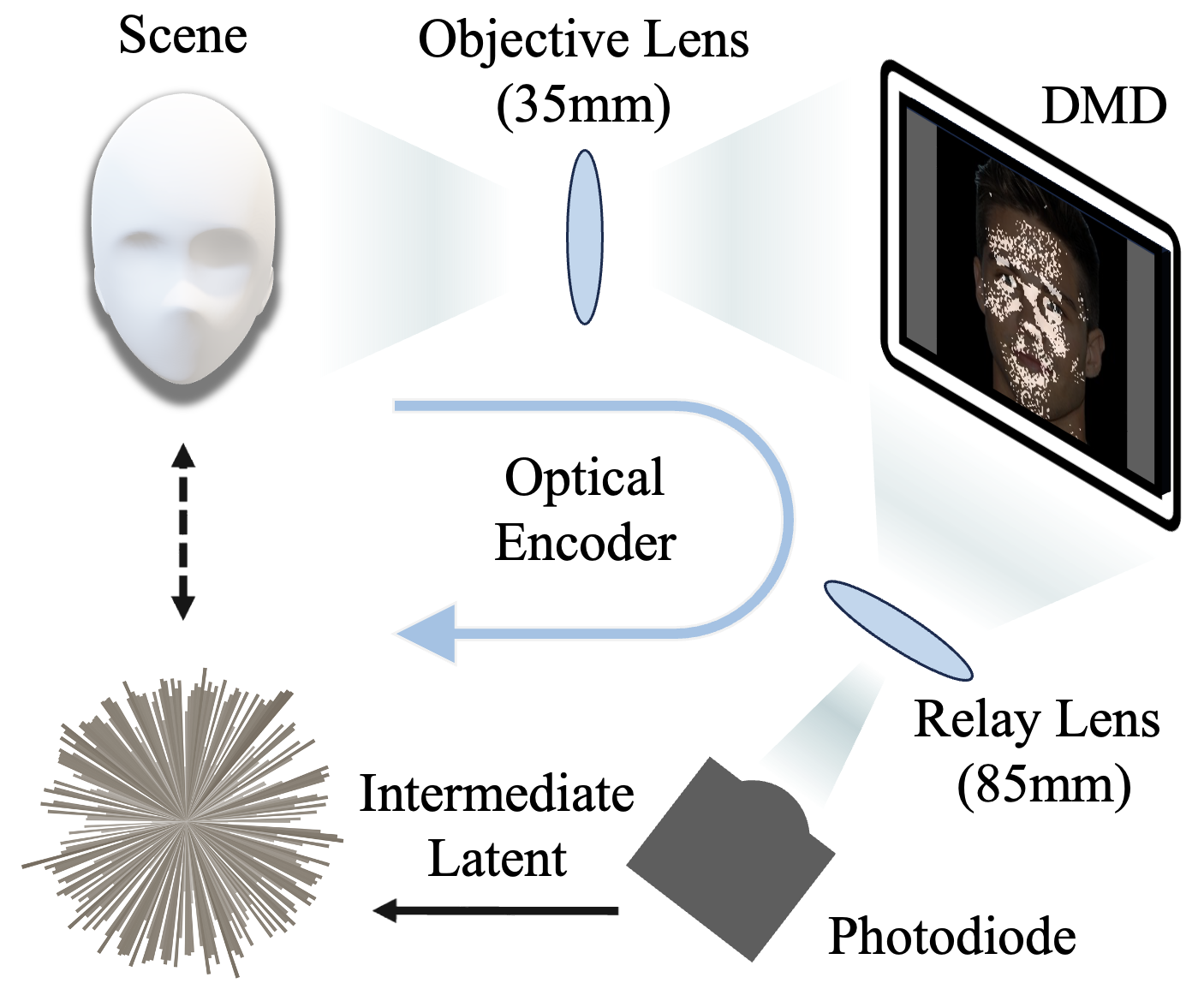}
    \caption{We illustrate one possible implementation of the Latent Space Imaging technique using a single-pixel framework. An objective lens focuses the image onto the Digital Micromirror Device (DMD), which is responsible for implementing the learned mask to spatially modulate the incoming signal. This is followed by a relay lens, which focuses the modulated signal onto a photodiode (SPD) responsible for integrating the signal. Utilizing time multiplexing, we can retrieve the necessary measurements.}
    \label{fig:setup}
\end{figure}

Realizing the LSI framework (as detailed in Fig.~\ref{fig:setup}) involves
implementing the optical encoder hardware along with the digital
encoder software module. Among a range of different alternatives to
realize this linear layer in real-world, we opt for the single-pixel
camera approach as the hardware framework for our prototype, since it
allows for programmable adjustments of the compression ratio for
experimentation. The Supplement discusses alternative options for
physical realization of the LSI concept.

A schematic of our setup is depicted in Fig.~\ref{fig:setup}. In this
system, a scene is projected onto a Digital Micromirror Device (DMD TI
DLP4500) using an objective lens. The image is then spatially
modulated by the optimized patterns. The modulated output is captured
by a relay lens, which focuses it onto a single-pixel detector
(ThorLabs PDA100A2). This detector acts as an integrator, measuring a
single value that represents the integrated intensity of the entire
modulated image. A computer controls and synchronizes the operation of
the DMD and the measurement signals, enabling the time-multiplexing of
all measurements that compose the compressed signal.

In the single-pixel camera implementation of the LSI framework,
$\optenc$ spatially modulates the incoming image $\img$ through matrix
multiplication (denoted by $\cdot$ in Eq.~\ref{eq:mapping}). Each
row of the optical encoder corresponds to one learned mask, and each
column represents one measurement. Each mask corresponds to a single
pixel, where its forms are learned for a specific domain/imaging
purpose.  Having the hardware in the loop during training involves
maintaining $\optenc$'s entries as positive, bounded between [0, 1],
and quantized. In order to maintain the proposed setup as minimal as
possible, we utilize binary optical masks - corresponding to
quantizing its values to 0 or 1. Such decision enables reaching high
frame-rate on light modulators like Digital Mirror Devices (DMDs).  

\begin{figure}[h!]
    \centering
    \includegraphics[width=0.4\textwidth]{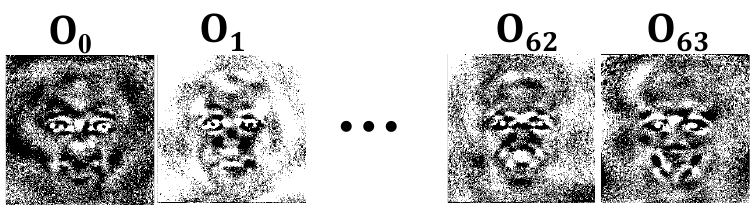} 
    \caption{ Displays optimized pixel forms for 1:1024 compression ratio.}
    \label{fig:physical_b}
\end{figure}

With this framework established, we further improve the system’s light
efficiency and sensitivity to subtle variations by enforcing a $1\%$
intensity difference between each optical mask. This distinction is
achieved by introducing an energy efficiency loss in the later stages
of training, fine-tuning the encoding process to ensure each pattern
is uniquely identifiable, thereby significantly enhancing the system’s
ability to effectively discriminate between different patterns. Fig.~\ref{fig:physical_b} illustrates that
the masks result in patterns highlighting the
system's focus on facial landmarks for specific domain encoding. Please
refer to the Supplement for details.

\section{Training Details}\label{sec:training}
\begin{figure*}[!ht]
    \centering
    \includegraphics[width=0.9\textwidth]{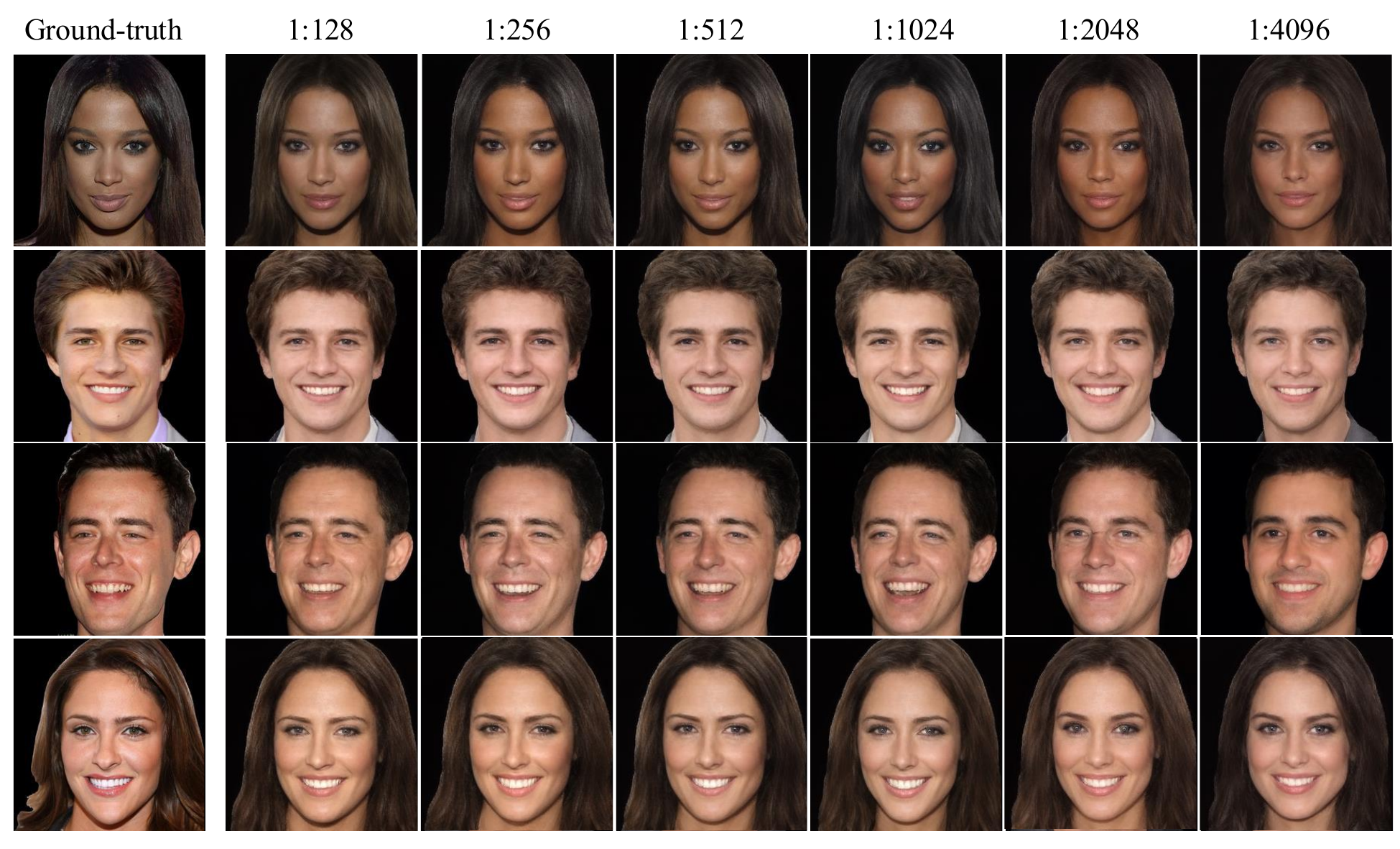} 
    \caption{Illustration of reconstructions achieved through the LSI pipeline with bounded and quantized pixels, using compression ratios from 1:128 to 1:4096 in the simulation setting.}
    \label{fig:result}
\end{figure*}

\textbf{Optical Encoder Optimization}  
In optimizing $\optenc$, we use the straight-through estimator (STE)~\cite{NIPS2016ste}, as in VQ-VAEs~\cite{oord2018neural, razavi2019generating, mentzer2024finite}. Forward passes quantize and constrain $\optenc$ values, while back-propagation allows gradients to flow through the non-quantized version, facilitating effective weight updates. \\ 
\textbf{Datasets}  
We utilize FFHQ~\cite{karras2019stylebased} and CelebAHQ~\cite{karras2017progressive}, reserving 2000 images from CelebAHQ for testing. Only CelebAHQ is used for downstream applications, where Segment Anything (SAM)~\cite{kirillov2023segany} focuses on faces, disregarding background content. The Supplement includes additional evaluations on the AFHQ dataset~\cite{choi2020starganv2}. \\
\textbf{Loss Functions}  
Alongside the latent~\ref{eq:minimize} and energy~\ref{sec:resphy} losses, we employ the identity loss ($\mathcal{L}_{id}$), calculating cosine distance between feature maps from ArcFace~\cite{deng2019arcface}. We also apply the $\ell_2$ norm ($\mathcal{L}_{l2}$) and DINO/LPIPs features as perceptual loss ($\mathcal{L}_{p}$)~\cite{zhou2024deformable,zhang2018unreasonable}. 

The Supplement provides full details on hyperparameters, optimizers, loss functions, as well as downstream application losses and additional training technicalities.

\section{Simulation Results}\label{sec4}

In our experiments we evaluate the LSI pipeline with simulated results as well as real-world measurements. We focus on the face domain understanding how the minimal imaging performs on essential facial tasks like reconstruction, identity recognition, attribute classification, segmentation, and landmark detection, analyzing the boundaries for each one of such tasks.

For quantitative results, we select two metrics for face identification: VGGFace~\cite{parkhi2015deep} and Dlib~\cite{kazemi2014one}. In both cases, pre-trained face recognition models are used to extract features from the original and reconstructed images. Cosine similarity is then computed to verify identity; a similarity score below the thresholds of 0.68 and 0.07, respectively, indicates a match. These thresholds are sourced from the DeepFace~\cite{deepface} library. Additionally, FID~\cite{heusel2017gans} is also computed to compare the distribution of features between the reconstructed and the original set. Traditional metrics, such as PSNR and SSIM, are useful for evaluating pixel-level accuracy but often fall short in fully capturing human visual perception~\cite{zhang2018unreasonable, ledig2017photo, blau2018perception}. Further assessment and an in-depth discussions of these metrics are provided in the Supplement.

\begin{table*}[ht]
    \centering
    \begin{tabular}{c *{6}{>{\centering\arraybackslash}p{1.6cm}}}
        \toprule
        Comp. Ratio & VGGFace$\uparrow$ & DLib$\uparrow$ & FID$\downarrow$ & Acc.$\uparrow$ & $F1$$\uparrow$ & $NME_{dg}$$\downarrow$ \\
        \midrule
        1:128 & 91.97\% & 92.74\% & 27.38 & 89.07\% & 70.00\% & 1.48 \\
        1:256 & 90.98\% & 92.68\% & 26.62 & 89.15\% & 70.94\% & 1.43 \\
        1:512 & 89.61\% & 91.67\% & 28.66 & 89.20\% & 70.25\% & 1.48 \\
        1:1024 & 81.12\% & 87.44\% & 28.79 & 88.74\% & 69.18\% & 1.52 \\
        1:2048 & 54.72\% & 77.77\% & 35.89 & 88.06\% & 65.81\% & 1.67 \\
        1:4096 & 27.22\% & 59.21\% & 46.18 & 86.44\% & 60.63\% & 2.01 \\
        1:8192 & N/A & N/A & N/A & 83.81\% & 53.77\% & 2.18 \\
        1:16384 & N/A & N/A & N/A & 81.75\% & 46.36\% & 2.47 \\
        \bottomrule
    \end{tabular}
    \caption{Latent Space Imaging is quantitatively evaluated for simulated results across multiple downstream tasks under highly compressed settings. For the image reconstruction task, validation is performed using both the VGGFace~\cite{arcface} and DLib~\cite{kazemi2014one} image recognition pipelines. Performance metrics include Fréchet Inception Distance (FID), classification accuracy, F1-mean score for segmentation tasks, and Normalized Mean Error (NME), normalized by the diagonal of the face bounding box. In cases of extreme compression ratios, while recognition metrics and reconstruction may not be applicable (N/A), downstream applications remain feasible.}
    \label{tab:results}
\end{table*}

\subsection{Image Reconstruction}

Full image reconstruction is the most demanding task for any reduced
sampling scheme, especially in the case of human faces, where the
human visual system is extremely sensitive to minute differences. 
Fig.~\ref{fig:result} demonstrates the robust identity preservation capabilities of our model when reconstructing images at a resolution of $256 \times 256$ across a range of measurement levels. Latent Space Imaging effectively captures essential features such as eye color, expression, facial shape, and skin tone with less than $0.4\%$ of the original pixel count. This capability is achieved by designing $\optenc$ to selectively retrieve relevant features from the target domain, enabling focused extraction of critical characteristics for facial identification while preserving landmarks embedded in the underlying distribution learned by the pre-trained generative model $\gen$. The model $\gen$ was trained exclusively on face images (FFHQ 256x256), which is the same training data used for all comparison methods. This targeted feature retrieval, along with compressing the target space into a flat latent structure, enables the reconstruction process not only to interpolate under-sampled data but also to generate outputs closely resembling the original, detailed facial images.

The compression ratios shown in Fig.~\ref{fig:result} correspond to
measurement levels of $512$, $256$, $128$, $64$, $32$, and $16$, with
quantitative assessments provided in
Table~\ref{tab:results}. Crucially, only $64$ measurements are already
sufficient for accurate recognition with ArcFace~\cite{arcface} and
Dlib~\cite{kazemi2014one}, with accuracy declining significantly below
$32$ samples. However, even for these extreme compression ratios,
critical facial features—such as facial structure, gaze direction, and
expression—are remarkably preserved. This underscores the advantage of
targeting the latent space of generative models over achieving
pixel-perfect accuracy. Beyond visualization, the linear boundaries
within the latent space offer significant benefits for downstream
applications, enabling effective feature recognition with minimal
adjustments.

\section{Prototype Results}

\begin{figure}[!ht]
    \centering
    \includegraphics[width=0.46\textwidth]{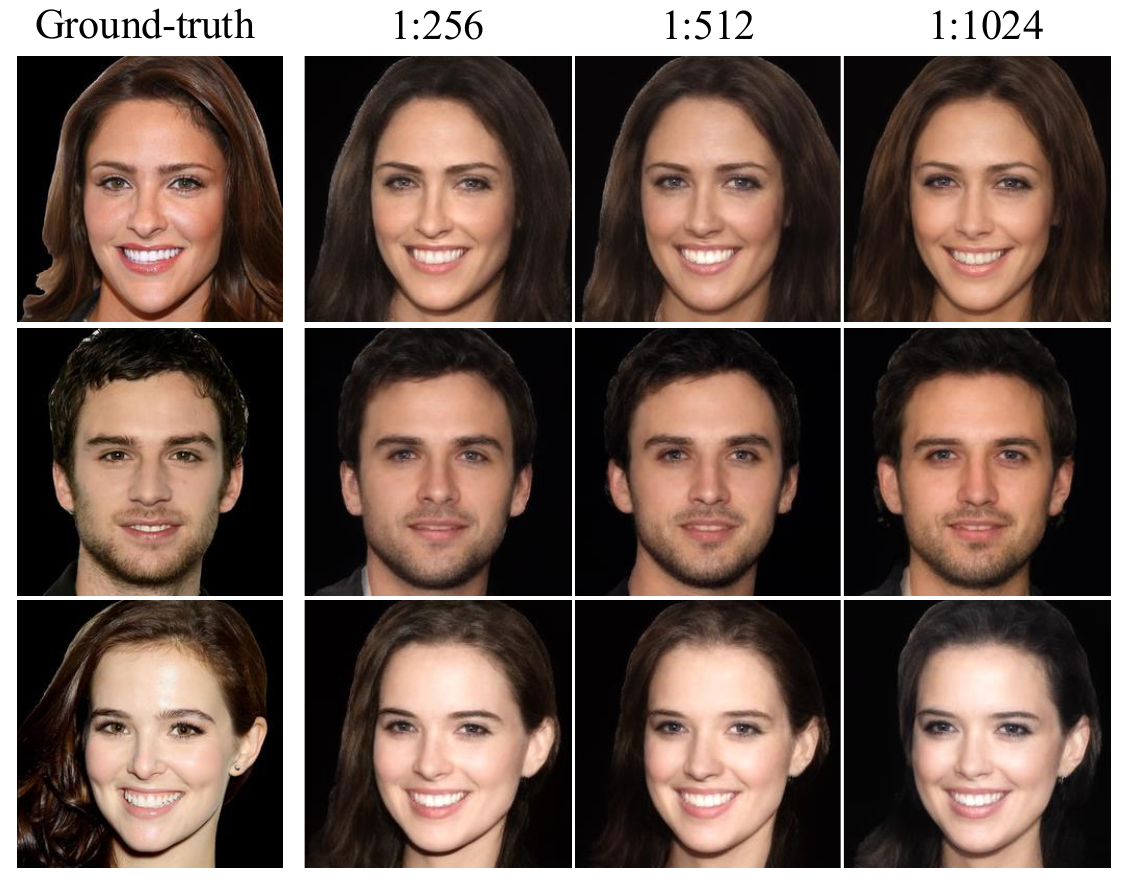} 
    \caption{Face reconstructions from the experimental setup across different compression ratios.}
    \label{fig:real_measurement}
\end{figure}
Real world experiments are conducted using our prototype and the pipeline described in Sec.~\ref{sec:resphy}. A set of 100 face images is randomly selected from CelebAHQ test set and displayed on a high-dynamic-range Eizo CG3145 monitor, then captured by the single-pixel setup. Fig.~\ref{fig:real_measurement} illustrates the reconstruction results across three different compression ratios. These same measurements are further utilized by the linear projectors to produce attributes classification, segmentation masks and landmark points for compression ratios of 1:512, 1:2048 and 1:8192 (see Figs.~\ref{fig:attributes},~\ref{fig:faceparsing} and~\ref{fig:landmarks}), corresponding to $128$ and $32$ scalars, respectively, thereby validating the feasibility of the LSI framework.

In the following, we demonstrate how our sensing matrix, optimized for the latent space, effectively supports a wide range of downstream computer vision tasks \emph{without} requiring full image reconstruction. All results presented are based on \emph{real-world experiments} conducted using the prototype described in Sec.~\ref{sec:resphy}.

Each application is evaluated on the CelebAHQ~\cite{karras2017progressive} dataset, following the architectures and feature projections detailed in Sec.~\ref{sec:semantics}. For attribute classification, we use the original 40 labels per image as ground truth. In landmark detection, we leverage FaRL~\cite{zheng2022general} as a state-of-the-art model to compute 68 landmark points, using the original image $\img$ as the reference. For segmentation tasks, we empirically found that generating masks using reconstructed faces $\imgo$ with FaRL as ground truth achieves superior accuracy, given the spatial precision needed for effective segmentation.

\subsection{Attributes}

\begin{figure}[!ht]
    \centering
    \includegraphics[width=0.4\textwidth]{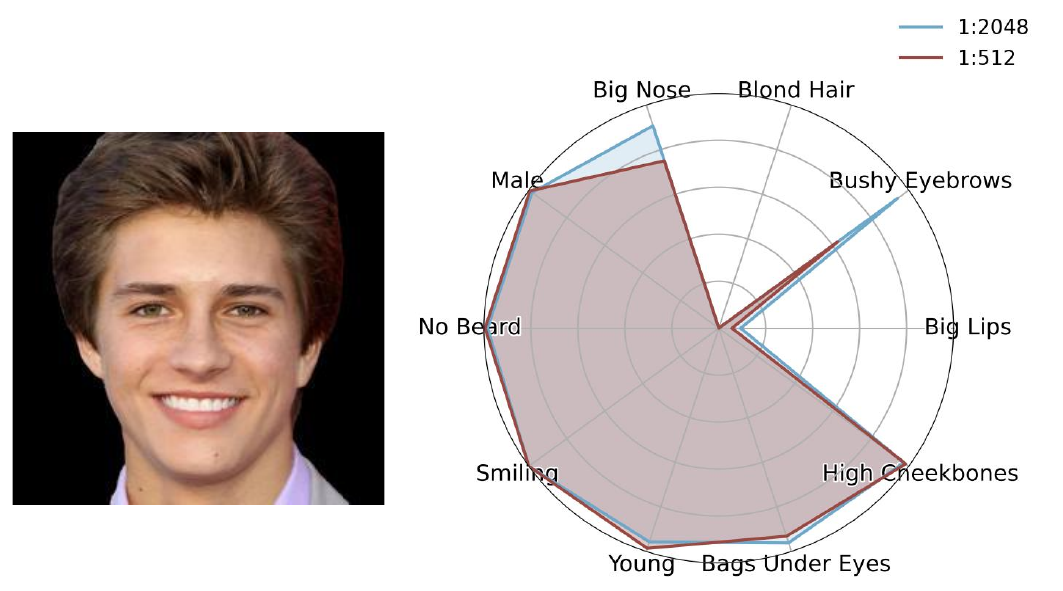} 
    \caption{The figure illustrates the sigmoid-transformed logits for
      compression ratios of 1:256 and 1:2048, overlaid for
      comparison. These correspond to 256 and 32 scalars,
      respectively.
      }
    \label{fig:attributes}
\end{figure}

The classification accuracy of 40 binary attributes is evaluated across the full range of compression ratios. Notably, we achieve $80\%$ precision even at a 1:16384 compression ratio, which corresponds to only 4 scalar measurements. Fig.~\ref{fig:attributes} visualizes the sigmoid-transformed logits after linearly projecting the latent space, $\proj_{A}(\latentplus)$. Despite the decline in similarity observed in Table~\ref{tab:results} and Fig.~\ref{fig:result}, the selected attribute ratios remain close to the original, offering an accurate representation of the original face.

\subsection{Facial Landmarks}

\begin{figure}[!ht]
    \centering    \includegraphics[width=0.46\textwidth]{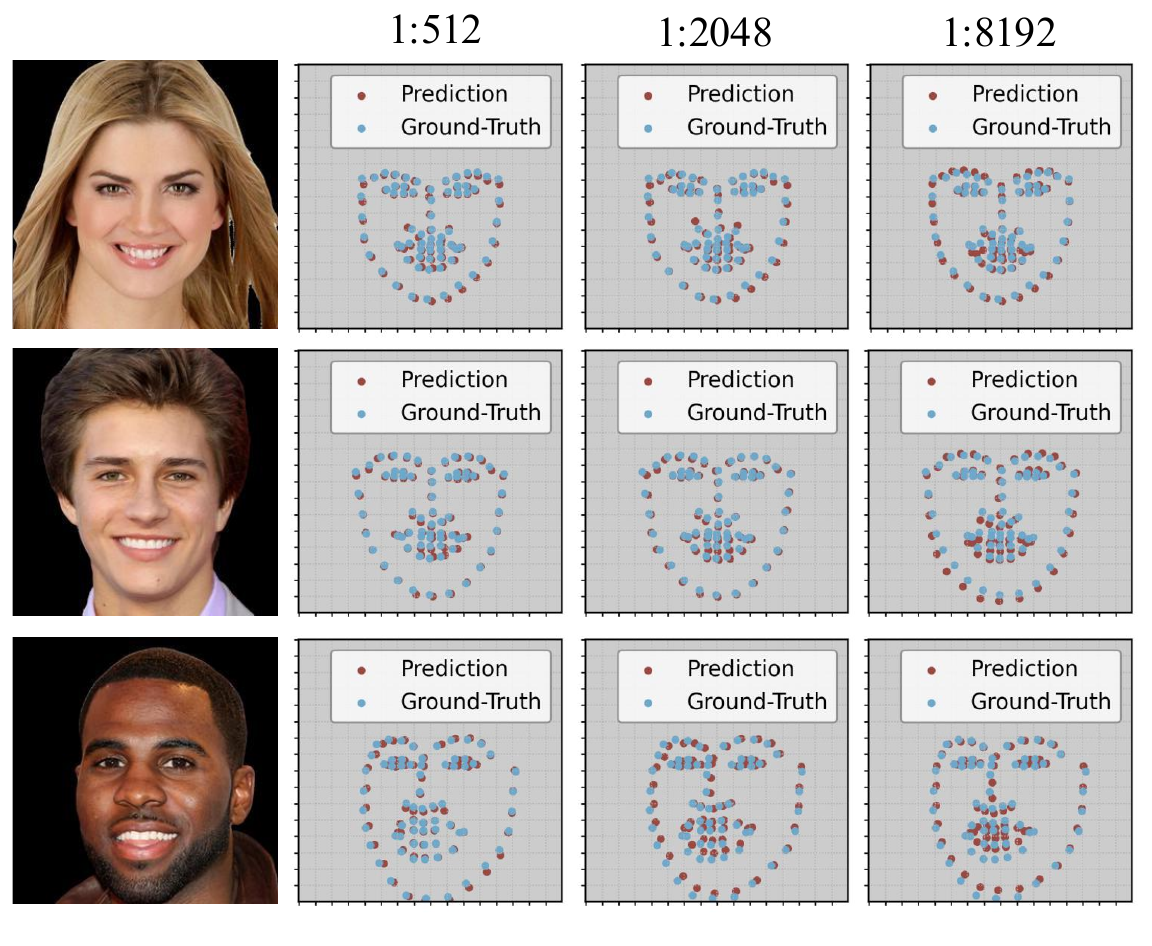} 
    \caption{Overlay of predicted and ground-truth positions for 68 facial landmarks across different compression ratios.}
    \label{fig:landmarks}
\end{figure}

Fig.~\ref{fig:landmarks} highlights landmark detection acquired by the linear projection $\proj_{L}(\latentplus)$ on the same two images across various compression levels, demonstrating high accuracy even in extremely compressed scenarios, such as 1:8192 (8 scalars), compared with the pseudo ground-truth ($gt$) generated by FaRL~\cite{zheng2022general}. The third row shows a non-frontal view of a face, yet the predictions remain close to the reference, indicating robustness.

The quantitative evaluation in Table~\ref{tab:results} uses the Normalized Mean Error (NME) metric (Eq.~\ref{eq:nme}), scaled by $\alpha = 100$ for visualization purposes. NME is calculated as the average Euclidean distance between predicted and ground-truth landmarks, normalized by the diagonal of the face bounding box ($d^{\text{dg}}$) to account for variations in face size.

\begin{equation}
\text{NME}_{\text{dg}} = \frac{\alpha}{N} \sum_{i=1}^{N} \frac{\| \proj_{L}(\latentplus)_{i} - {gt}_i \|_2}{d^{\text{dg}}_{i}}.
\label{eq:nme}
\end{equation}

\subsection{Face Segmentation}

The segmentation task is performed using the FaRL model, trained on the CelebAHQ dataset, which includes 18 classes. During our linear projection, \( \proj_{S} \), we aim to learn the entirety of these classes. However, due to the limitations of our generative model, reproducing high-frequency details and unique objects are challenging; for instance, accessories like earrings are not well segmented. Despite this, for general facial features such as nose, hair, face, mouth, ears, and eyebrows, the linear projection of $\gen$ internal features performs well, even under highly compressive settings.

\begin{figure}[!ht]
    \centering
    \includegraphics[width=0.46\textwidth]{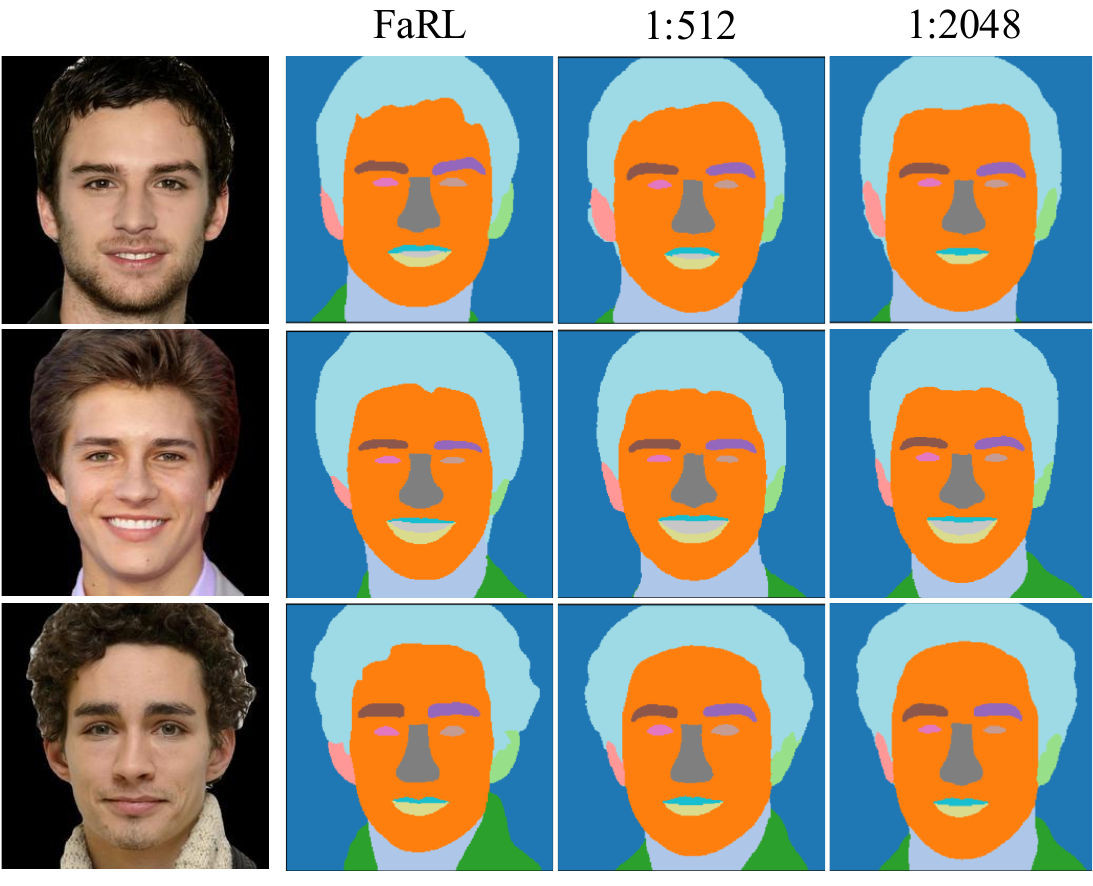} 
    \caption{Illustration of face segmentation for two compression levels (1:512 and 1:2048) as well as the pseudo ground-truth for extract from the original image utilizing the FaRL model.}
    \label{fig:faceparsing}
\end{figure}

Fig.~\ref{fig:faceparsing} illustrates segmentation results across a
range of different hair styles and different facial hair, showing that
the segmentation masks remain consistent despite a four times increase
in compression ratio.
Additionally, Table~\ref{tab:results} highlights the stability of the
F1 score across different compression ratios, demonstrating reliable
segmentation performance.

Extended results and discussions for all compression ratios, downstream tasks, evaluation metrics, detailed descriptions of the physical setup are provided in the Supplement.

\section{Comparison with Alternative Approaches}
\label{sec:comparison}

\begin{figure}[!ht]
    \centering
    \includegraphics[width=0.47\textwidth]{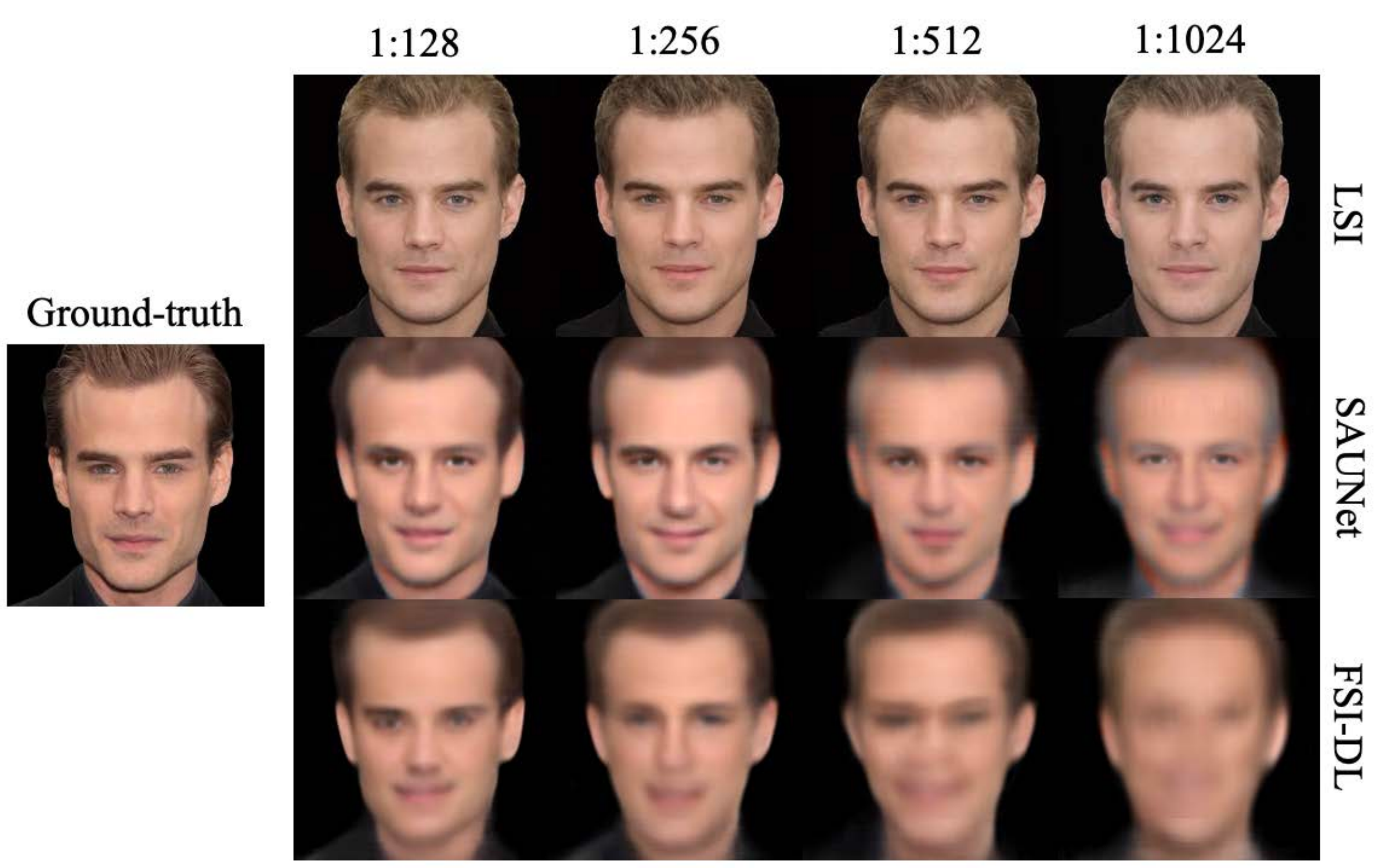} 
    \caption{Qualitative comparison between different methods.}
    \label{fig:comparison}
\end{figure}

Although our approach diverges from current strategies in compressed
imaging, we selected notable models for comparison. Specifically, we
included representative models from block-based sensing matrices~\cite{song2023optimization,tcsnet},
unfolding networks~\cite{wang2023saunet}, traditional single-pixel methods with deep
learning-based reconstruction~\cite{ferri2010differential,Wang:22}, Fourier pattern single-pixel
methods~\cite{zhang2015single}, and optimization-based methods using generative models as
priors~\cite{jalal2020robust}. Table~\ref{tab:compa} and Fig.~\ref{fig:comparison} illustrate
the comparison quantitative and qualitative simulated results for deep
learning single-pixel using Fourier features~\cite{zhang2015single}
and unfolding networks~\cite{wang2023saunet}. For fair comparison, the
sensing matrix for the unfolding network is 1 bit and 8 bit quantized
which is required for a plausible physical
implementation. Quantization training follows the same strategy adopted by LSI.
Detailed methodologies and all mentioned comparisons are elaborated in the Supplement.

Fig.~\ref{fig:comparison} highlights the details of our model which contrasts with traditional compressive sensing strategies, where limited measurements lead to reconstruction algorithms over-smoothing the images, resulting in the loss of critical identity features such as the shape of eyes, facial hair, and other distinguishing landmarks. Table~\ref{tab:compa} also highlights that even with better quantization deep unfolding methods still outperformed by LSI.

\begin{table}[ht]
    \centering
    \begin{tabular}{c *{3}{>{\centering\arraybackslash}p{1.6cm}}}
        \toprule
        & VGGFace$\uparrow$ & Dlib$\uparrow$ & FID$\downarrow$ \\
        \midrule
        FSI-DL & 3.30\% & 13.15\% & 123.5 \\
        SAUNet  & 17.84\% & 50.12\% & 75.53 \\
        SAUNet\textsuperscript{\textdagger} & 39.47\% & 65.97\% & 65.79 \\
        LSI & {\bf90.98}\% & {\bf92.68}\% & {\bf26.62} \\
        \bottomrule
    \end{tabular}
    \caption{Quantitative comparison between different compressed imaging methods for a 1:256 ratio. SAUNet\textsuperscript{\textdagger} represents
      a version retrained on 8 bits quantized sensing matrix, while the SAUNet was 1 bit quantized, similar to LSI and FSI-DL.}
    \label{tab:compa}
\end{table}

\vspace{-0.2cm}
\section{Discussion and Conclusion}\label{sec:future}

Compared to other imaging techniques, Latent Space Imaging
offers substantial benefits in domain-specific scenarios with tight
requirements for hardware complexity, memory and bandwidth
requirements, or frame rate. As mentioned in the introduction, LSI is
strongly inspired by biological vision, particularly the human visual
system. The relationship between LSI and biological vision is
discussed in more detail in the Supplement.

LSI inherits the advantages of a well-trained generative model
specialized in a specific domain.  Consequently, its encoding learns
to represent the most meaningful and essential information within this
domain. This capability enables high-quality measurements with high
compression ratios. However, it also inherently possesses the
limitations of the underlying generative model and the inversion
network.  Therefore, the model may fail to reconstruct very specific
features not present during training, such as hats, glasses, makeup,
or other distinct image details, and may struggle to generalize to a
multi-domain approach.

Nonetheless, we believe that LSI is a promising new paradigm for
domain-specific imaging systems, including privacy-preserving cameras
that nonetheless excel at a specific imaging task. LSI is also a
promising avenue for imaging systems ultra-low hardware requirements
(both in terms of optical/electronic complexity and compute power),
and/or ultra-high frame rate imaging. These applications will require
alternative hardware implementations compared to our single-pixel
camera prototype, which was chosen primarily due to easy
reconfiguarbility. Several concepts for such alternative hardware
designs are discussed in the Supplement.

{
    \small
    \bibliographystyle{ieeenat_fullname}
    \bibliography{main}
}

\clearpage
\setcounter{section}{0} 
\renewcommand{\thesection}{\Alph{section}} 
\setcounter{figure}{0} 
\renewcommand{\thefigure}{\Alph{section}.\arabic{figure}} 

\setcounter{table}{0} 
\renewcommand{\thetable}{\Alph{section}.\arabic{table}} 

\maketitlesupplementary

This supplementary document consists of three
parts: \begin{enumerate*}[I.)]\item A more in-depth discussion of
fundamental principles of Latent Space Imaging, including a discussion
of the relationship to biological vision, and of alternative optical
encoding schemes. \item Additional implementation details of the
current system, including the hardware prototype and training
details. \item Additional experimental results and ablation
studies.\end{enumerate*}

\section{Part I: Foundations of LSI}

\subsection{Relationship to Biological Vision}
As mentioned in the main text, LSI is strongly inspired by biological
vision, in particular the human visual system (HVS). Specifically, in
the HVS, photoreceptors sense the spatial distribution of light, and
retinal ganglion cells (RGCs) encode these spatial distributions into
a compressed latent space transmitted over the optic nerve.  Finally
the visual cortex decodes and processes the information. The
compression ratio between the number of photoreceptors and the number
of axions in the optic nerve is about 100:1, and is enabled by the
inherently non-linear processing of the RGCs. 

LSI can be seen as analogous to this process: in our single pixel
camera setup, the pixels of the spatial light modulator take on the
role of the photoreceptors -- they define the spatial resolution of
the sensed image. The optical codes shown on the SLM, combined with
the digital encoder take on the role of the RGCs, and both encode and
compress this information, while the StyleGANXL takes on the role of
the visual cortex.

The encoding module is therefore composed of a combination of a
(linear) optical computing part and a non-linear digital encoder. The
digital encoder is needed since optical
computing~\cite{wetzstein2020inference} is in practice limited to
linear operators. Critically, the most compressed representation is
the intermediate latent space between the optical and the digital
encoder, which enables highly compressed, low-bandwidth sampling
hardware. We show that, with this two stage encoding, very high
compression ratios can be achieved for the specific domain of facial
images: ratios of 1:100 to 1:1000 for full image reconstruction, and
even higher ratios for simpler tasks such as landmark detection or
segmentation.

The principle of LSI can be applied to the latent space of any
generative model, however the achievable compression ratios will
depend on the complexity of the domain. To match the performance of
the HVS on general vision tasks with a compression ration of 1:1000, we
will need both a general AI generative model as well as a way to
optically implement non-linear encodings.

\subsection{Alternative Physical Implementations}\label{sec:physical}
In our prototype, we showcase the physical implementation using
binary structural masking with a Digital Micromirror Device (DMD),
where the light source is a monitor projecting the images. The SPI
framework is an established platform for prototyping and testing different
compressed imaging approaches, since the DMD acts as a programmable
encoding element, making it easy to experimentally validate different
code patterns and compression ratios.

However, SPI also has several downsides that limit its practicality
for real-world imaging systems, including its form factor, and, most
notably, its difficulty in dealing with moving scenes. However, the
recent efforts in optical computing~\cite{wetzstein2020inference} have
resulted in a range of alternative options for implementing the
required linear optical encoding layer.

For example diffractive optical elements have been used to implement
both convolutional~\cite{chang2018hybrid} and fully connected linear
layers~\cite{lin2018all}, and meta-surfaces have been demonstrated for
shift variant convolution kernels~\cite{wei2023spatially}. Due to the
linear nature of light, all these methods are currently limited to
linear operators. However, as we show with the SPI setup, linear
optical encoders are sufficient for achieving high compression rates
when combined with a small digital encoder. As such, any of the
recently proposed optical compute frameworks could be used instead of
the SPI framework, although typically this would require ``freezing''
the optical encoder into hardware. Nonetheless, for special purpose
imaging systems, such an approach presents a practical avenue for
developing real-world, snapshot-capable LSI systems.

In the longer term, LSI may be able to benefit from ongoing research
efforts to develop non-linear optical computing hardware, for example
based on non-linear optical materials, quantum effects, polarization,
or other non-linear effects. Eventually, this may enable more powerful
optical encoders, thereby reducing or even eliminating the need for
the digital encoder module.

\section{Part II: Implementation Details}

\subsection{Digital Encoder Architecture} \label{sup:dea}

\begin{figure*}[!ht]
  \centering
  {\includegraphics[width=1.5\columnwidth]{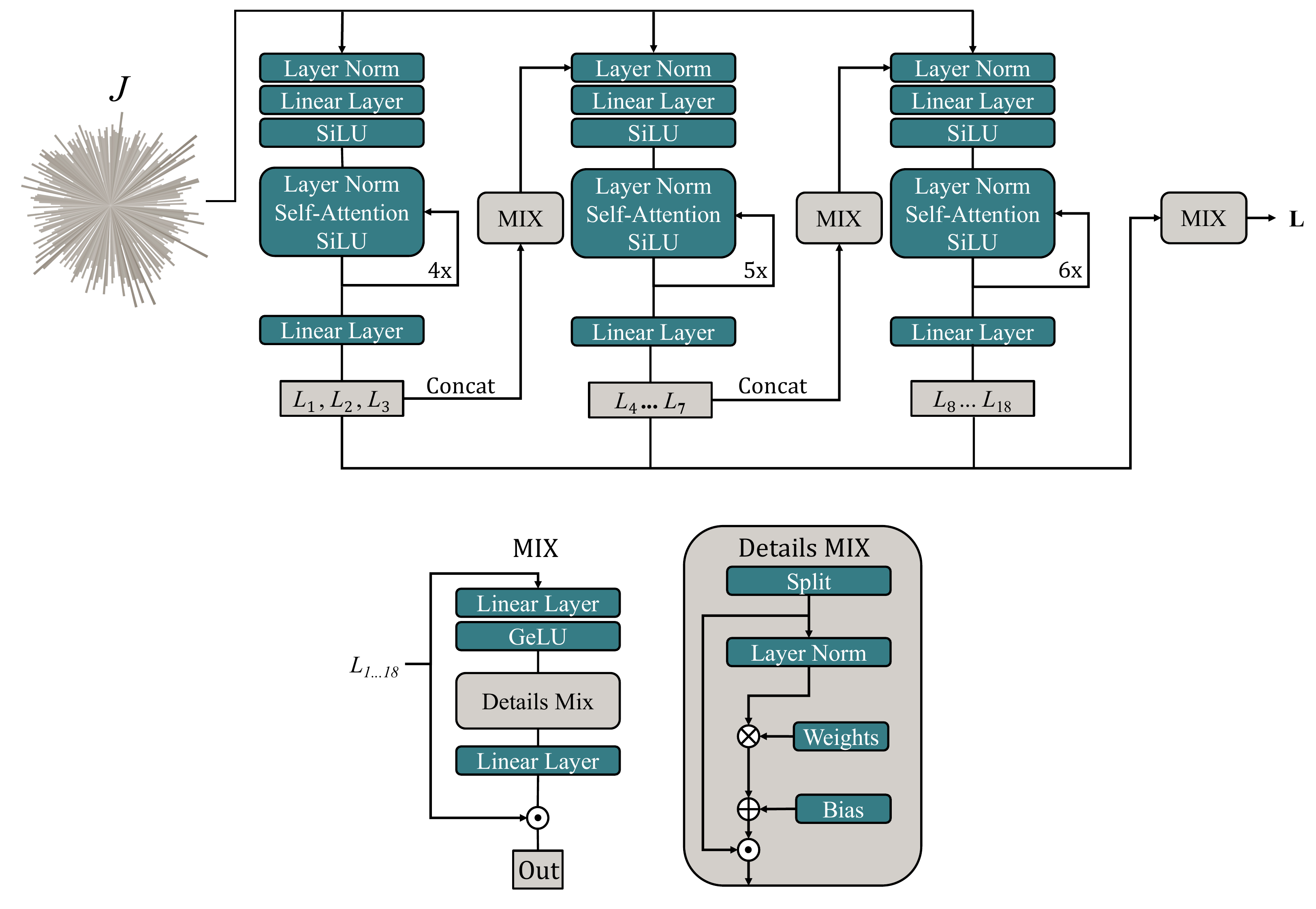}}
  
      \caption{(Top) Overview of all the blocks involved in the digital encoder architecture. The three steps capture coarse, middle, and fine details, with each step depending on the output of the previous one. (Bottom) Illustration of the MIX process. The latent vector is statically projected to capture interactions among different levels of detail.}
      \label{fig:net}
\end{figure*}

After acquiring the measurements $\imgc$ from the co-optimized optical encoder $\optenc$, we employ a digital encoder network $\digenc$. The objective of this model is to match the intermediate latent space produced by the measurements with the actual latent able to reproduce the image of interest $\img$.


The original StyleGANXL implementation receives a random noise vector and utilizes a mapping network to project this to different levels of details, result in  $\latentplus \in \mathbb{R}^{512 \times \gendimensiond}$, $\gendimensiond$ may vary depending the desired image resolution that the generative models was trained to generate. In the case of face images, $\gendimensiond$ is equal to 18, finally we want $\optenc$ to map $\imgc \in \mathbb{R}^{\dimension} $ to $\latentplus \in \mathbb{R}^{512 \times 18}$. $\dimension$ corresponds to our measurement count, therefore, going from 512 to 4.

As depicted by Fig.~\ref{fig:net}, the intermediate latent goes first through 3 different set of blocks from left to right. The first one produces the first 3 elements of $\latentplus$ out of 18, each one of them only take into account the initial $\imgc$. The middle block, besides $\imgc$, it also takes the concatenation of $\textit{L}_{1},\textit{L}_{2}$ and $\textit{L}_{3}$ through the MIX module, as input, outputting $\textit{L}_{4}$ to $\textit{L}_{7}$. Finally, the last block follows the same idea, it concatenates the output of the middle one and pass through another MIX block, the output of this is summed together with the original measurement and it passes through the block to finally result in $\textit{L}_{8}$ to $\textit{L}_{18}$. The idea here is to make the optical-aware inversion network match the multi-level structure of StyleGANXL, following a coarse-to-fine approach.

In the end, the latents from all detail levels pass through the MIX block. The purpose of the MIX blocks is to facilitate interactions and learn a weighted mixture across multiple detail levels. The model incorporates the spatial gating unit proposed in \cite{liu2021pay}. Within the MIX block (see Fig.\ref{fig:net}), the features are projected to a higher dimension, after which the resulting tensor ${x}$ is split into two parts, ${u}$ and ${v}$, along the channel dimension. This division allows for separate processing paths within the block. ${v}$ is normalized and linearly projected using a weights matrix and bias vector to capture interactions in a static manner, as the weights matrix remains unchanged after training because it does not depend on the input. The final element-wise multiplication of ${u}$ and the projected ${v}$ modulates the information flow, controlling which parts of ${u}$ are allowed to pass through—similar to a gating mechanism. Finally, this gated output is merged with the input through a residual connection, preserving the original information and enhancing gradient flow during back-propagation.

\subsection{Loss Functions and Training Details} \label{sup:losst}

In this work, we trained the optical $\optenc$ and digital $\digenc$ encoders while keeping the generative model, StyleGANXL, frozen. To jointly optimize these components, we trained an off-the-shelf inversion network pre-trained to invert images to the latent space of a specific domain~\cite{richardson2021encoding}. This network is used to compute a latent similarity loss function ($\mathcal{L}_{lat}$), making the output of the $\digenc$ similar to the latent space approximated by this model. This process is described in Equation 3 of the main text.

Additionally, we aim to match the image quality at the pixel level and enforce identity similarity. For this task, we used the identity loss ($\mathcal{L}_{id}$), which computes the cosine distance between feature maps extracted by ArcFace~\cite{deng2019arcface}, a facial recognition network. We also utilized the $\ell_2$ norm ($\mathcal{L}_{l2}$) and DINO/LPIPs features loss($\mathcal{L}_{p}$)~\cite{zhou2024deformable,zhang2018unreasonable} to enforce pixel-wise and perceptual similarities, respectively. The details of DINO feature extraction are provided in \cite{zhou2024deformable}.

The total loss function during training can be summarized as follows:

\begin{equation}
\mathcal{L}_{total} = \lambda_{lat}\mathcal{L}_{lat} + \lambda_{id}\mathcal{L}_{id} + \lambda_{p}\mathcal{L}_{p} + \lambda_{l2}\mathcal{L}_{l2}
\label{eq:loss}
\end{equation}

After convergence, we added an additional loss term ($\lambda_{energy}\mathcal{L}_{energy}$) to account for the intensity diversity among the masks.

\begin{equation}
\label{eq:energy}
\begin{matrix}
\mathcal{L}_{energy} =  \frac{1}{\dimension}\sum_{j}^{\dimension}  \left | \sum_{i}^{mn} \mathcal{\optenc}_{i,j} - \epsilon_{j} \right |
\end{matrix}
\end{equation}

\begin{figure}[!htt]
    \centering
    \includegraphics[width=0.4\textwidth]{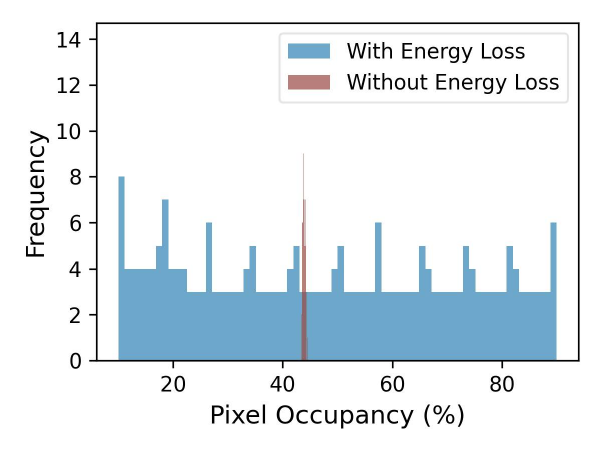} 
    \caption{Shows the histogram of pixel occupancy, comparing scenarios where the energy loss function is and is not utilized. The use of energy modeling enhances the dynamic range of the measurements captured by the single sensor.}
    \label{fig:physical_a}
\end{figure}

Here, $\epsilon$ represents the ground-truth energy level, heuristically designed to correspond to a certain percentage of pixels set to one. This percentage increases incrementally from $10\%$ to $90\%$, with a step of $1\%$. For strong compression from 1:1024 to 1:16384 the min and max energy boundaries move closer to $50\%$ accordingly. The patterns are shuffled to avoid any undesired structure.   As demonstrated in Fig.~\ref{fig:physical_a}, with the energy loss applied, the pixel occupancy histogram showcases a broad spectrum of occupancies, reflecting a high
diversity of patterns. In contrast, without this modification, the
patterns are more uniform in their occupancy levels, exhibiting less
variability. 

To optimize $\optenc$ and $\digenc$, we employ distinct optimizers: Lion~\cite{chen2024symbolic} for $\optenc$ and Ranger (combination of Lookahead~\cite{NEURIPS2019_90fd4f88} and the Rectified Adam~\cite{liu2019variance}) for $\digenc$, each with a learning rate of $10^{-4}$. The weighting coefficients $\lambda$ for each loss function are set at $1, 0.5, 0.8,$ and $1$, in the respective order that they appear in  Equation~\ref{eq:loss}. Additionally, for the energy loss component, we set $\lambda$ to 3. Regarding batch sizes, we typically use 32 for face datasets. However, for the AFHQ~\cite{choi2020starganv2} dataset, as discussed in Sec.~\ref{sup:addres}, the batch size is reduced to 8 due to the limited number of images available. For all experiments we utilize a single NVIDIA A100.

\subsection{Optical Encoder Optimization Details} 

In the $\optenc$ optimization process, in order to deal with the quantized patterns, we employ the straight-through estimator (STE)~\cite{NIPS2016ste}, a technique also favored by VQ-VAEs~\cite{oord2018neural, razavi2019generating, mentzer2024finite}. During the forward pass, values are quantized and constrained before computing $\imgc$. However, during back-propagation, gradients are allowed to flow through the non-quantized version, facilitating weight updates. Importantly, we ensure that the entries of $\optenc$ remain positive and are bounded between $[0, 1]$. Although using complementary patterns~\cite{xu2024compressive} allows for the inclusion of negative values, empirical experiments have shown that this approach does not yield significant improvements and necessitates double the measurements.

Initially, masks are generated from a uniform distribution and are then binarized through a quantization process. This process results in binary patterns that achieve an even total counting of 0s and 1s, theoretically maximizing bit entropy and steering the system towards an optimal solution~\cite{lin2022siman}. This distribution presumes the sensor is sensitive enough to discern very fine differences, a demanding prerequisite as each pattern will have similar total intensity. This issue was addressed using the energy loss mechanism.

\subsection{Experimental setup} \label{sup:expset}

\begin{figure}
    \centering
    \includegraphics[width=0.8\linewidth]{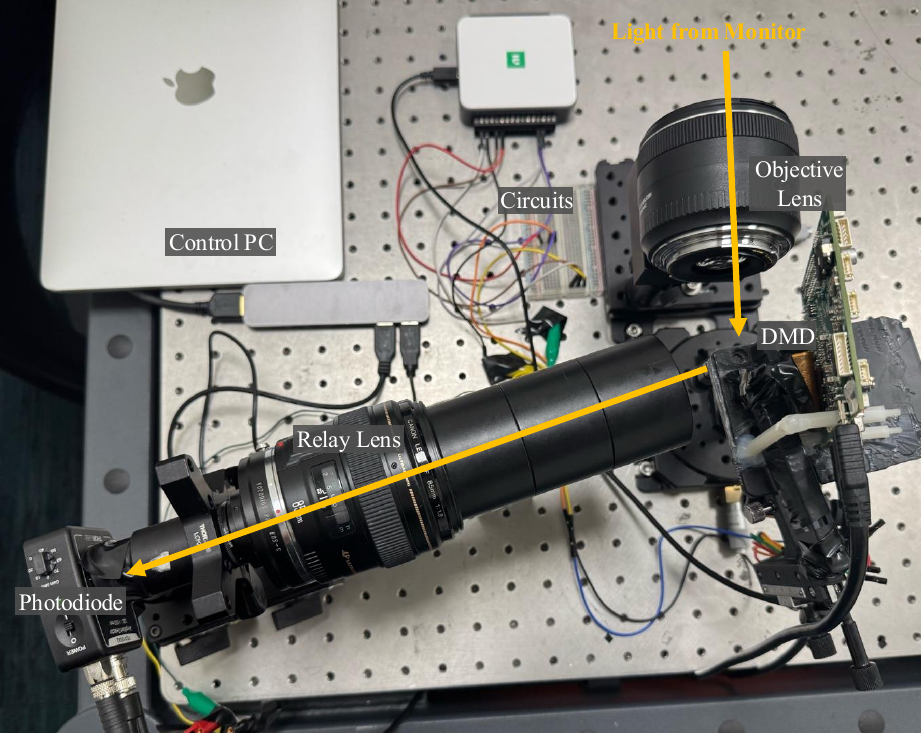}
    \caption{Experiment setup. An objective lens collects light from monitor and relays it to the DMD, which modulates the light with reflecting patterns. A relay lens directs the modulated light to the photodiode, and the PC records the measurements. Yellow arrows denote the light path. }
    \label{fig:setup_real_photo}
\end{figure}

Our experimental setup utilizes a single-pixel imaging configuration as follows: Face images are displayed on a high-dynamic-range Eizo CG3145 monitor. These images are then focused onto the DMD of the Texas Instruments DLPLCR4500EVM evaluation board using a Canon EF 35mm f/2 IS USM objective lens. We interpolate our learned $256 \times 256$ patterns to optimally fill the largest possible area, adjusting the scale to compensate for the diamond-shaped micromirror grid of the DMD. This adjustment ensures full utilization of the height within a central square region. The DMD spatially modulates the image by reflecting patterns according to an optimized encoding matrix ($\optenc$). The resultant patterned light is captured by a photodiode (PDA100A2) through a Canon EF 85mm f/1.8 USM relay lens. An analog-to-digital converter (NI 6001) converts the received analog signals into digital form. The digital data are then received by a connected PC, which is also responsible for synchronizing all components involved in the experimental setup.

To enhance the congruence between our measurements and simulations, we employed a white image to determine a global scaling deviation and applied a correction factor. This approach effectively reduces the gap between simulation and measurement, yet some inherent challenges persist due to non-linear behaviors, minor deviations in DMD reflection angles, and sensor noise, which complicate the precise replication of simulated conditions. However, leveraging the controlled environment of our experimental setup, we are able to systematically gather real-world training data and fine-tune $\digenc$.

The fine-tuning process involves selecting a subset of 200 training images and measuring their outputs using our configured setup. We employ the loss functions specified in Equation~\ref{eq:loss}, while reducing the learning rate to $10^{-5}$, to optimize performance and accuracy in real-world applications. 

\section{Part III: Additional Results} \label{sup:addres}

We present additional LSI results encompassing a diverse set of images, along with the range of compression levels.

Figures~\ref{fig:img1},~\ref{fig:img2}, and~\ref{fig:img3} demonstrate a consistent retention of facial details, particularly as compression increases, highlighting the effectiveness of our approach at higher compression rates. By targeting the latent space, our method addresses the over-smoothing often observed with aggressive sub-sampling strategies, preserving textures and producing realistic reconstructions instead of flat, featureless outputs. Notably, key facial features such as eyebrow shape, beards, and smiles are well reconstructed, emphasizing the persistence of facial expressions. In contrast, competitor models fail to achieve similar results.

\begin{figure*}[!ht]
  \centering
  \includegraphics[width=1.\linewidth]{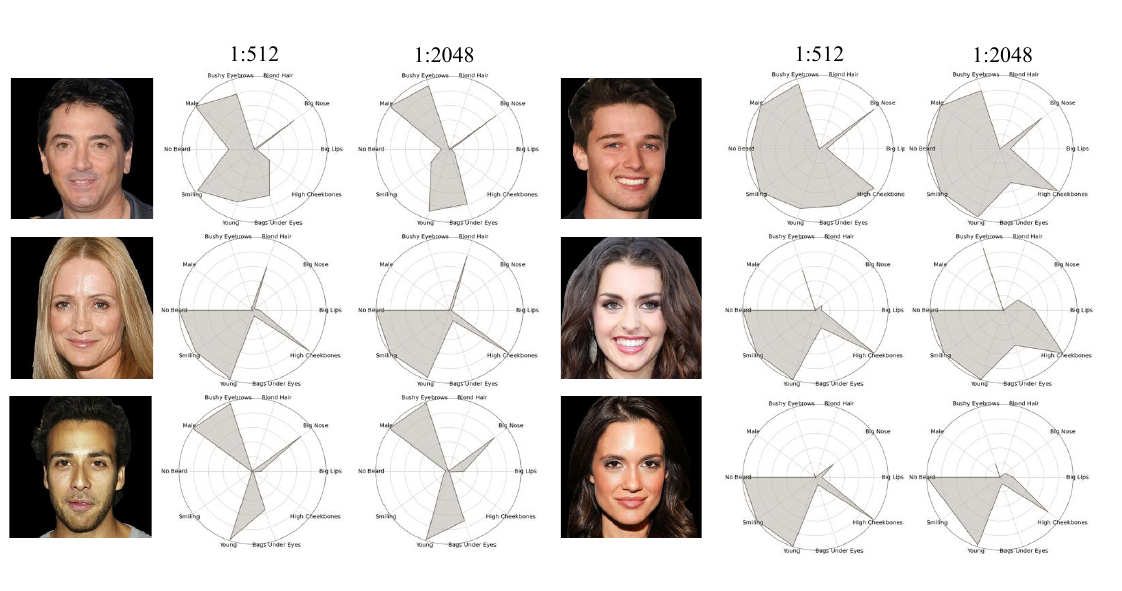}
      \caption{Additional attribute visualizations from the experimental validation are presented. The leftmost column displays the original faces as shown on the monitor, followed by reconstructed images at compression ratios of 1:512 and 1:2048, which correspond to 128 and 32 measurements, respectively.}
  \label{fig:real_att}
\end{figure*}

\begin{figure*}
    \centering
    \includegraphics[width=0.9\linewidth]{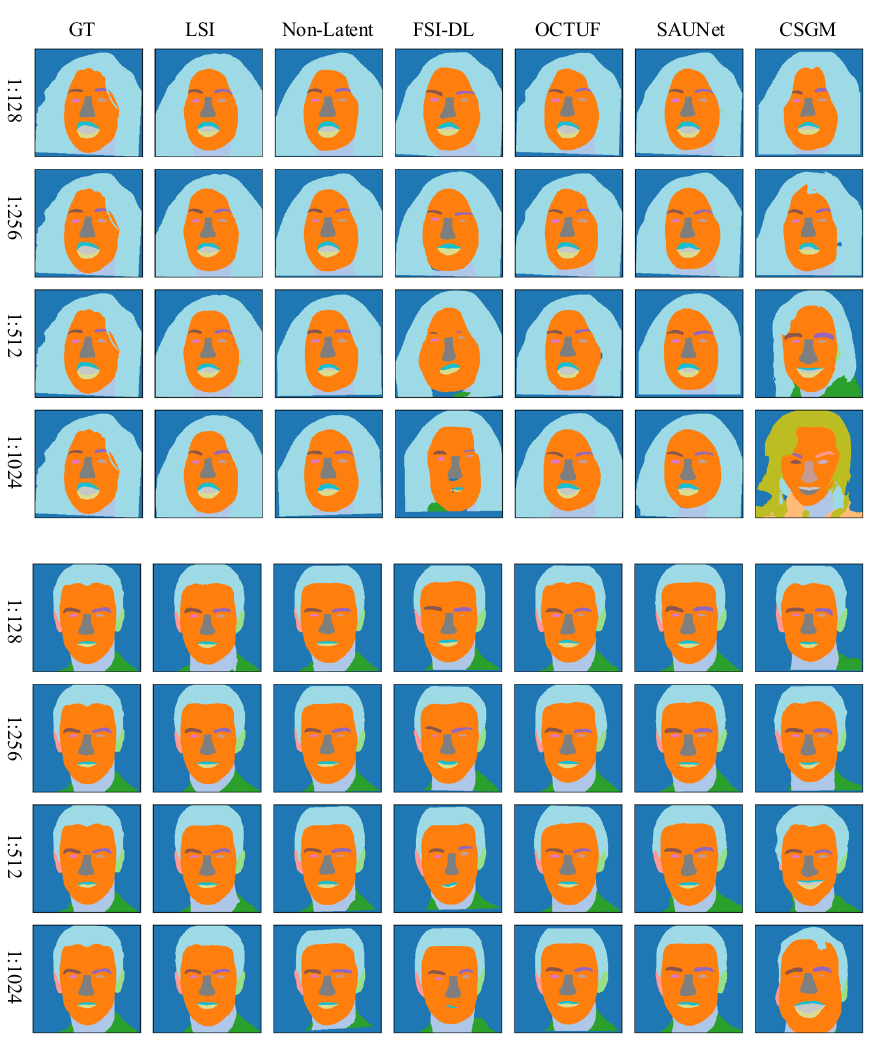}
    \caption{Facial segmentation comparison among different methods.}
    \label{fig:seg1}
\end{figure*}

\begin{figure*}
    \centering
    \includegraphics[width=0.9\linewidth]{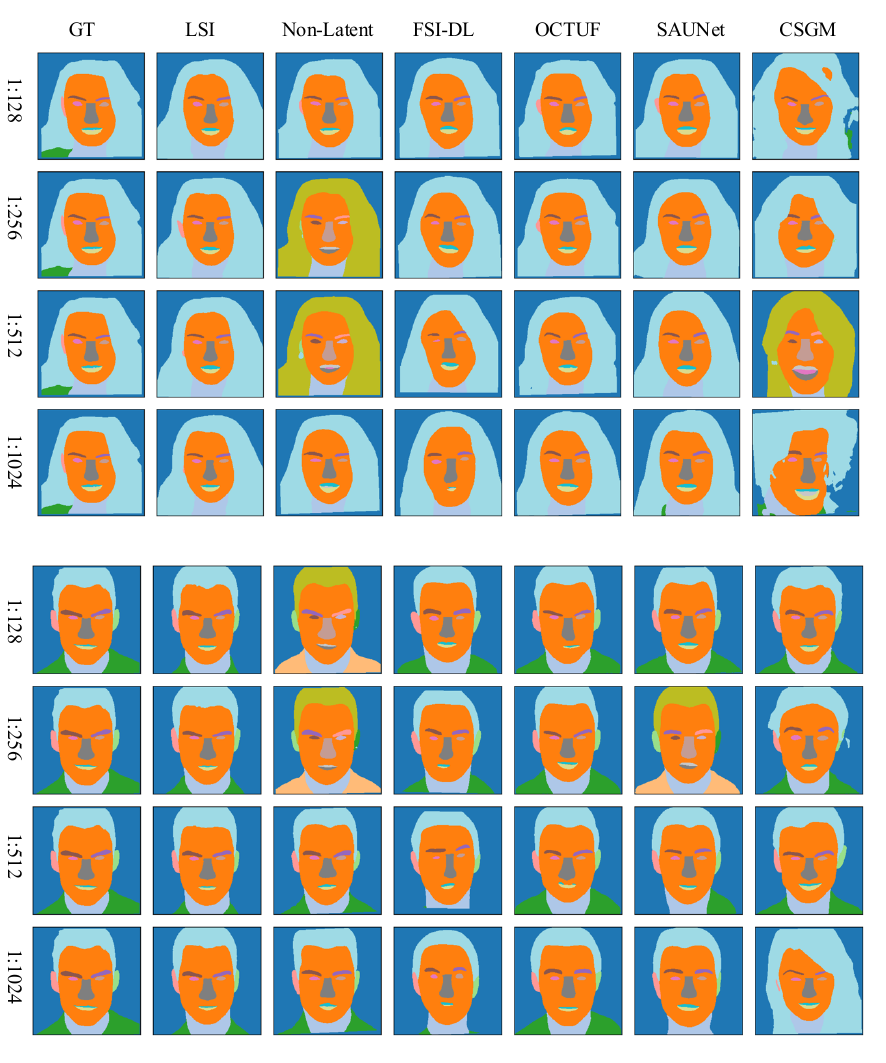}
    \caption{Facial segmentation comparison among different methods.}
    \label{fig:seg2}
\end{figure*}

\begin{figure*}
    \centering
    \includegraphics[width=0.9\linewidth]{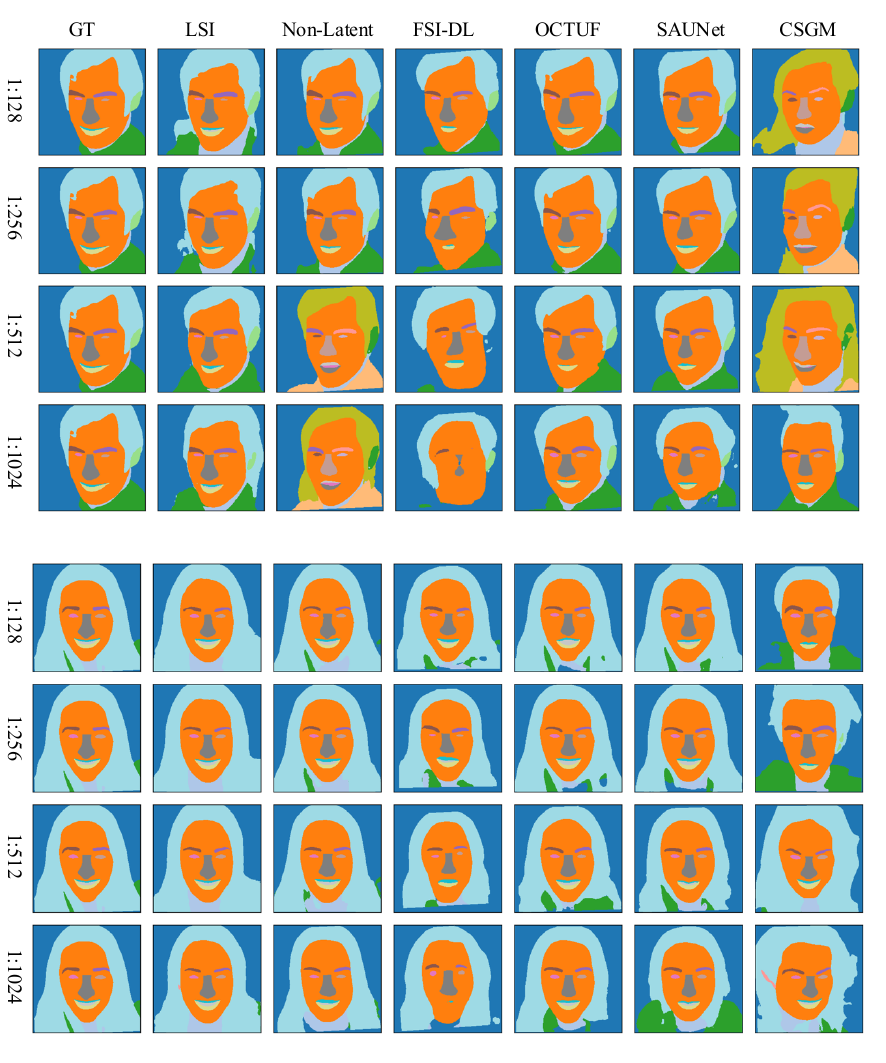}
    \caption{Facial segmentation comparison among different methods.}
    \label{fig:seg3}
\end{figure*}

\begin{figure*}
    \centering
    \includegraphics[width=0.9\linewidth]{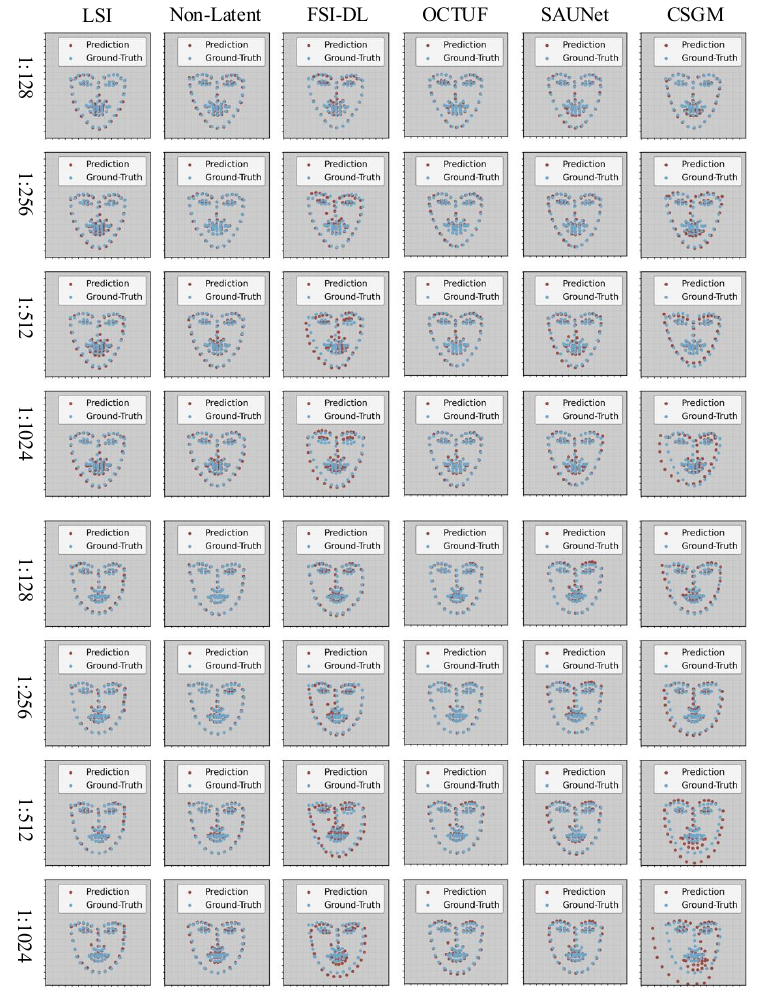}
    \caption{Landmarks detection comparison among different methods.}
    \label{fig:land1}
\end{figure*}

\begin{figure*}
    \centering
    \includegraphics[width=0.9\linewidth]{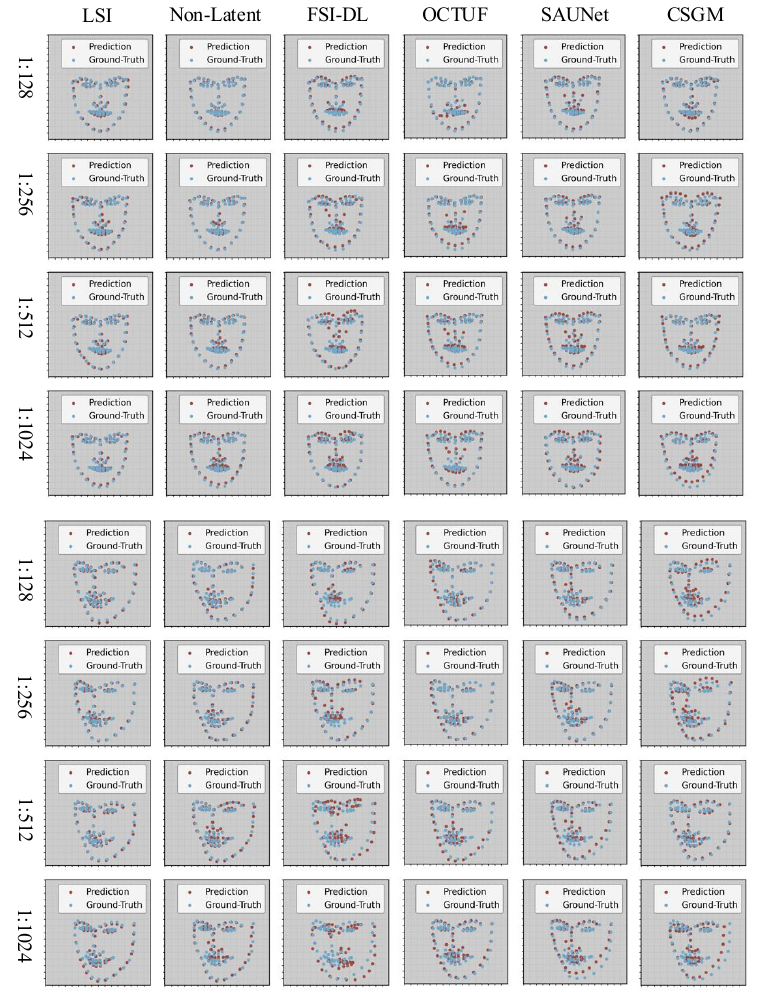}
    \caption{Landmarks detection comparison among different methods.}
    \label{fig:land2}
\end{figure*}

\begin{figure*}
    \centering
    \includegraphics[width=0.9\linewidth]{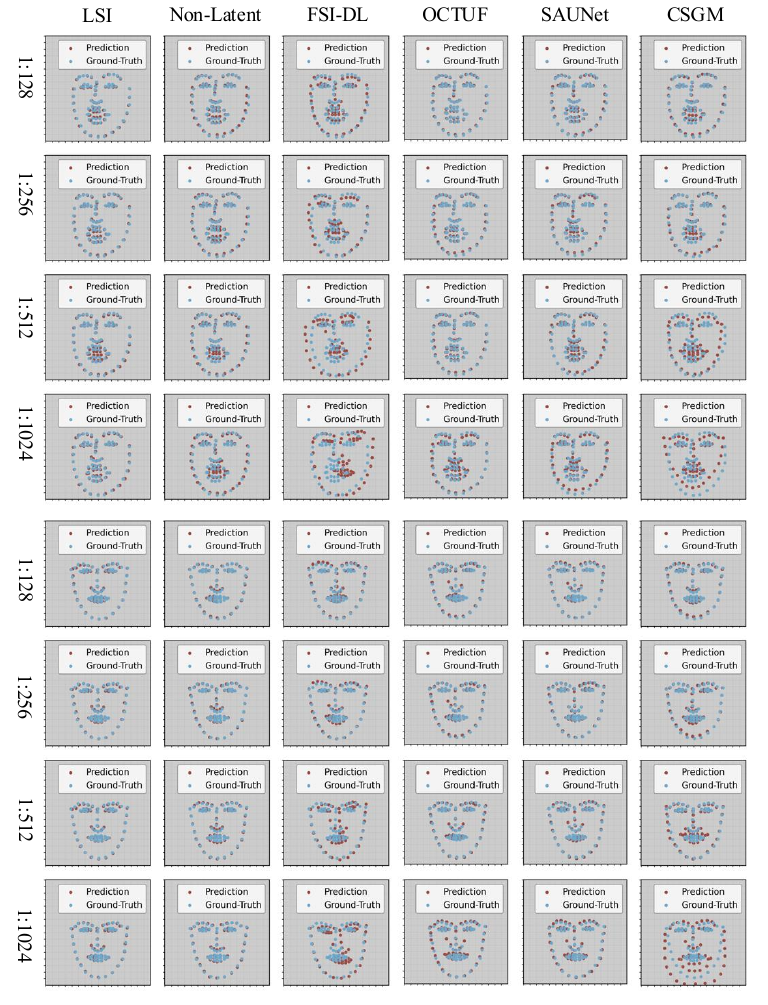}
    \caption{Landmarks detection comparison among different methods.}
    \label{fig:land3}
\end{figure*}

For the downstream tasks, Fig.~\ref{fig:real_att} presents additional results from the \textbf{experimental setup} for attribute classification. To further evaluate and enhance the assessment of our method, we include an extensive set of qualitative comparisons (see Figs.~\ref{fig:seg1},~\ref{fig:seg2},~\ref{fig:seg3},~\ref{fig:land1},~\ref{fig:land2}, and~\ref{fig:land3}) against other methods. These comparisons focus on simulated results for face segmentation and landmark detection.

Table~\ref{tab:results_sup} quantitatively summarizes our results, highlighting the overall superiority of our method in reconstruction and downstream tasks, particularly in high-compression setups.

\begin{figure}[!ht]
  \centering
  \includegraphics[width=0.9\columnwidth]{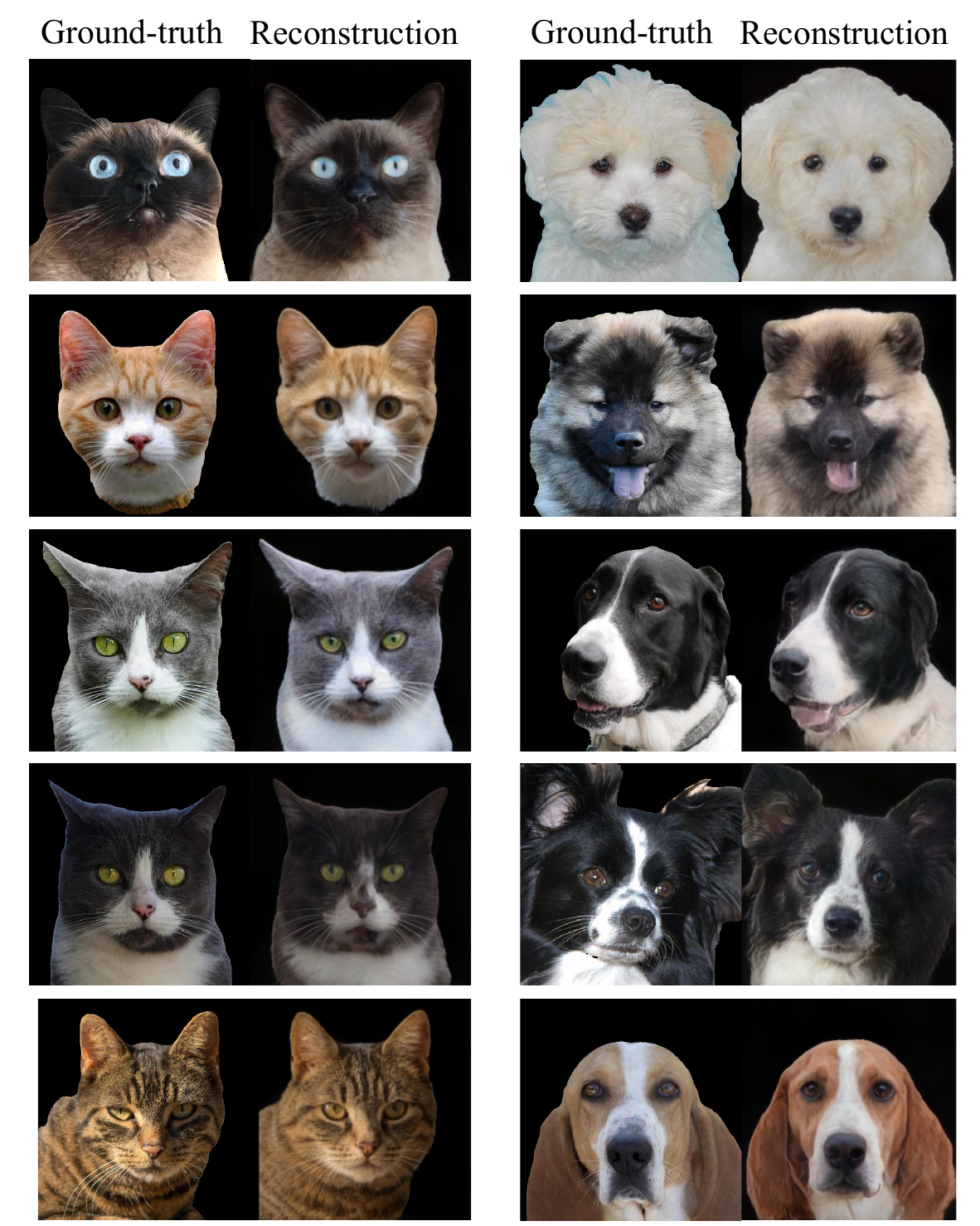}
      \caption{Illustrates another set of reconstructions from our simulated results, highlighting the versatility of our methods across various domains, such as cats and dogs. For this dataset we utilized StyleGAN2 instead XL, also showing the versatility of our method utilizing different generative models.}
  \label{fig:animals}
\end{figure}

Additional domains were explored in Fig.~\ref{fig:animals}, we utilize our simulated pipeline with 512 (1:128) measurements to reconstruct cats and dogs images from AFHQ dataset~\cite{choi2020starganv2}. These datasets imposed additional challenge because they are very small compared with FFHQ, with only 5000 training images. However, LSI is capable to faithfully reconstruct such domain pictures with a correspondingly trained encoder.

\begin{figure}[!ht]
  \centering
  {\includegraphics[width=0.6\columnwidth]{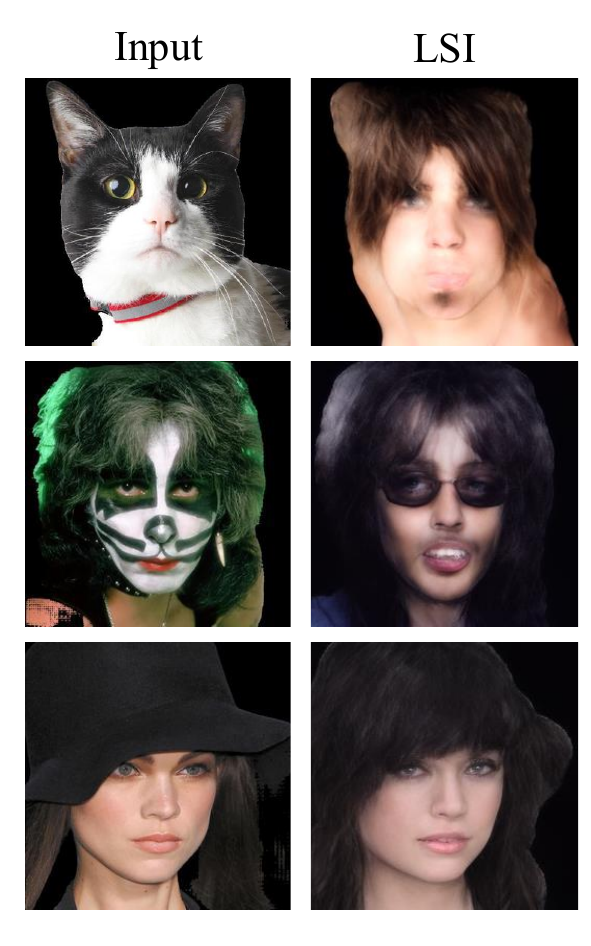}}
      \caption{Illustrates the conditions where LSI fails to reproduce the scene since the target scene is out of distribution of the generative model.}
  \label{fig:fail}
\end{figure}

\subsection{Out-of-Distribution Cases and Limitations}

As our focus is on faces, and our underlying generative model is primarily trained on frontal-centered faces under typical daily lighting conditions, significant deviations from these conditions are not well-modeled and often result in noticeable hallucinations. Fig.~\ref{fig:fail} illustrates several out-of-domain scenarios. For instance, providing a non-facial input, such as a cat, to the face generative model results in a hallucinated face that aligns with the characteristics of the given input signal. Similarly, accessories like hats or heavy makeup cannot be accurately modeled, as such data lies outside the model's training distribution.

\subsection{Qualitative and Quantitative Comparisons}

We evaluated a diverse set of representative models encompassing various approaches, including block-based sensing matrices, unfolding networks, single-pixel imaging techniques, and deep learning-driven single-pixel frameworks. Additionally, we incorporated optimization-based methods that leverage generative models as priors. To provide a broader perspective, we introduced a variant of our pipeline, referred to as \textbf{non-latent}, which replaces the generative model and latent representation with a state-of-the-art reconstruction network and a conventional signal recovery method. Further details on this model are provided below.
\\

\noindent \textbf{Preliminary notes on comparisons:}
\\
\begin{itemize}
    \item All methods were retrained on the same dataset and evaluated on the same testing dataset as LSI.
    \item SAUNet, FSI-DL, and Non-latent LSI utilize a quantized (binary) sensing matrix, similar to LSI. In contrast, the sensing matrices of other methods can represent any float value, giving them an (unrealistic!) advantage.
    \item SAUNet, OCTUF~\cite{song2023optimization}, and TCS-Net~\cite{tcsnet} are trained to reproduce luminance only, as proposed in their original papers. During inference, chrominance components were added using ground-truth values. Other methods, however, consider color images, modeling them either implicitly (via latent space) or explicitly (via demosaicking).
    \item For comparison purposes, downstream applications for all competitors were evaluated using FaRL~\cite{zheng2022general}, based on their reconstructed images, as they lack the inherent capability to perform these tasks. Because of low reconstruction quality, some competitor methods may fail the downstream tasks, we discard such data points only for the specific method.  
    Notably, for LSI, these results were achieved through a simple linear projection, entirely eliminating the need for a separate, complex model tailored to each specific task.
\end{itemize}

\begin{table*}[ht]
    \centering
    \begin{tabular}{l r *{6}{>{\raggedleft\arraybackslash}p{1.6cm}}}
        \toprule
        Method & Comp. Ratio & VGGFace$\uparrow$ & DLib$\uparrow$ & FID$\downarrow$ & Acc.$\uparrow$ & $F1$$\uparrow$ & $NME_{dg}$$\downarrow$ \\
        \midrule
        \multirow{4}{*}{LSI} & 1:128 & 91.97\% & 92.74\% & \textbf{27.38} & \textbf{89.07\%} & 70.00\% & 1.48 \\
                             & 1:256 & \textbf{90.98\%} & \textbf{92.68\%} & \textbf{26.62} & \textbf{89.15\%} & 70.94\% & 1.43 \\
                             & 1:512 & \textbf{89.61\%} & \textbf{91.67\%} & \textbf{28.66} & \textbf{89.20\%} & \textbf{70.25\%} & \textbf{1.48} \\
                             & 1:1024 & \textbf{81.12\%} & \textbf{87.44\%} & \textbf{28.79} & \textbf{88.74\%} & \textbf{69.18\%} & \textbf{1.52} \\
        \midrule
        \multirow{4}{*}{SAUNet} & 1:128 & 39.47\% & 64.12\% & 58.96 & 83.05\% & 74.15\% & 1.36 \\
                                & 1:256 & 17.84\% & 50.12\% & 75.53 & 82.66\% & 71.32\% & 1.59 \\
                                & 1:512 & 5.52\% & 31.42\% & 104.23 & 81.08\% & 66.79\% & 1.99 \\
                                & 1:1024 & 2.86\% & 16.59\% & 107.35 & 79.98\% & 62.37\% & 2.20 \\
        \midrule
        \multirow{4}{*}{FSI-DL} & 1:128 & 15.28\% & 32.60\% & 107.48 & 81.66\% & 62.20\% & 2.00 \\
                                & 1:256 & 3.30\% & 13.16\% & 118.79  & 79.43\% & 53.38\% & 3.20 \\
                                & 1:512 & 1.82\% & 1.93\% & 134.40 & 76.72\% & 41.45\% & 5.06 \\
                                & 1:1024 & 0.85\% & 2.05\% & 173.71 & 75.96\% & 34.95\% & 6.06 \\
        \midrule
        \multirow{4}{*}{OCTUF} & 1:128 & 54.99\% & 63.54\% & 70.83 & 83.93\% & 75.78\% & 1.31 \\
                               & 1:256 & 59.27\% & 73.09\% & 77.16 & 83.99\% & \textbf{74.78\%} & \textbf{1.26} \\
                               & 1:512 & 16.63\% & 43.31\% & 96.17 & 82.33\% & 69.92\% & 1.73 \\
                               & 1:1024 & 5.83\% & 26.62\% & 99.85 & 81.13\% & 66.18\% & 1.99 \\
        \midrule
        \multirow{4}{*}{TCS-Net} & 1:128 & 25.15\% & 49.69\% & 119.42 & 82.21\% & 66.00\% & 1.71 \\
                             & 1:256 & 1.32\% & 5.50\% & 248.60 & 77.71\% & 52.11\% & 3.31 \\
                             & 1:512 & 1.07\% & 2.62\% & 296.48 & 75.19\% & N/A & 7.43 \\
                             & 1:1024 & 0.11\% & 1.62\% & 333.34 & 74.05\% & N/A & 15.07 \\
        \midrule
        \multirow{4}{*}{CSGM} & 1:128 & 20.04\% & 51.85\% & 44.46 & 82.17\% & 62.20\% & 1.89 \\
                              & 1:256 & 13.66\% & 39.16\% & 46.30 & 81.28\% & 56.96\% & 2.36 \\
                              & 1:512 & 5.87\% & 21.96\% & 53.12 & 78.68\% & 46.07\% & 3.29 \\
                              & 1:1024 & 2.00\% & 10.23\% & 63.70 & 76.16\% & 38.31\% & 4.15 \\
        \midrule
        \multirow{4}{*}{Non-Latent} & 1:128 & \textbf{94.22\%} & \textbf{94.84\%} & 72.96 & 84.47\% & \textbf{77.47\%} & \textbf{1.07} \\
                                    & 1:256 & 87.52\% & 91.49\% & 75.62 & 83.71\% & 74.59\% & 1.26 \\
                                    & 1:512 & 71.82\% & 82.98\% & 81.87 & 83.08\% & 70.10\% & 1.53 \\
                                    & 1:1024 & 41.47\% & 68.22\% & 89.04 & 82.09\% & 64.73\% & 1.90 \\
        \bottomrule
    \end{tabular}
    \caption{Latent Space Imaging (LSI) and competing methods were quantitatively evaluated on simulated results across various downstream tasks, with compression ratios ranging from 1:128 to 1:1024. The results emphasize the superior performance of LSI, particularly in highly compressed scenarios. For the image reconstruction task, evaluations were conducted using the VGGFace~\cite{arcface} and DLib~\cite{kazemi2014one} image recognition pipelines. Metrics assessed include Fréchet Inception Distance (FID), classification accuracy, F1-mean score for segmentation, and Normalized Mean Error (NME), normalized by the diagonal of the face bounding box. Bold numbers highlight the best model for each metric and setting.}
    \label{tab:results_sup}
\end{table*}

\paragraph{AuSamNet and FSI-DL~\cite{Huang23}}

Both are Fourier basis methods with deep learning reconstruction algorithms; the first optimizes the mask, and the second utilizes a fixed heuristically designed circular mask. The patterns $P_{\phi}$ are generated using the ideal proposed by \cite{zhang2015single}

\begin{equation}
P_{\phi}(x, y; f_x, f_y) = a + b \cos(2\pi f_x x + 2\pi f_y y + \phi),
\end{equation}

where \((x,y)\) represents the 2D Cartesian coordinates in the scene, \(a\) symbolizes the average intensity distribution, \(b\) stands for the amplitude of the Fourier basis pattern, and \((f_x, f_y)\) indicates the non-zero spatial frequency points within the optimized mask. Furthermore, \(\phi\) denotes the initial phase, adopting three steps phase shifting of \(0\), \(2\pi/3\), and \(4\pi/3\).

This work also utilizes a color filter array (CFA) and demosaic process to reduce the number of measurements necessary to retrieve color images, computing the Fourier transform of the CFA and multiplying it with the mask.

\begin{figure*}
    \centering
    \includegraphics[width=1.0\linewidth]{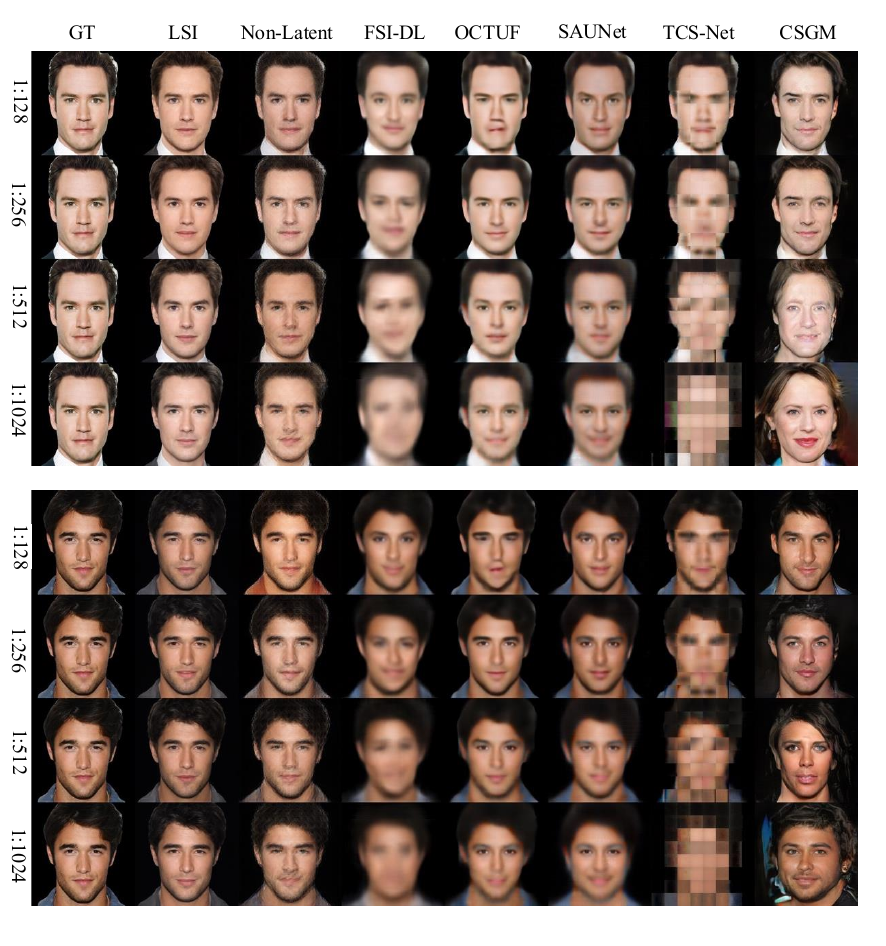}
    \caption{Different methods compared for facial reconstruction quality.}
    \label{fig:img1}
\end{figure*}

\begin{figure*}
    \centering
    \includegraphics[width=1.0\linewidth]{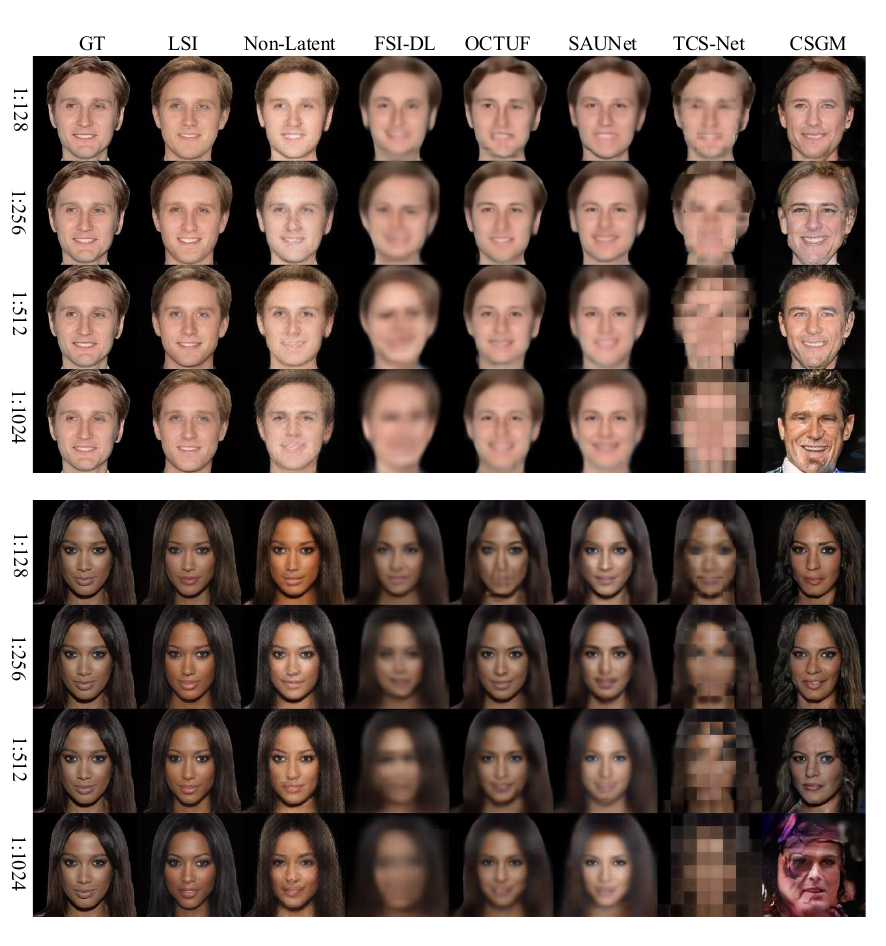}
    \caption{Different methods compared for facial reconstruction quality.}
    \label{fig:img2}
\end{figure*}

\begin{figure*}
    \centering
    \includegraphics[width=1.0\linewidth]{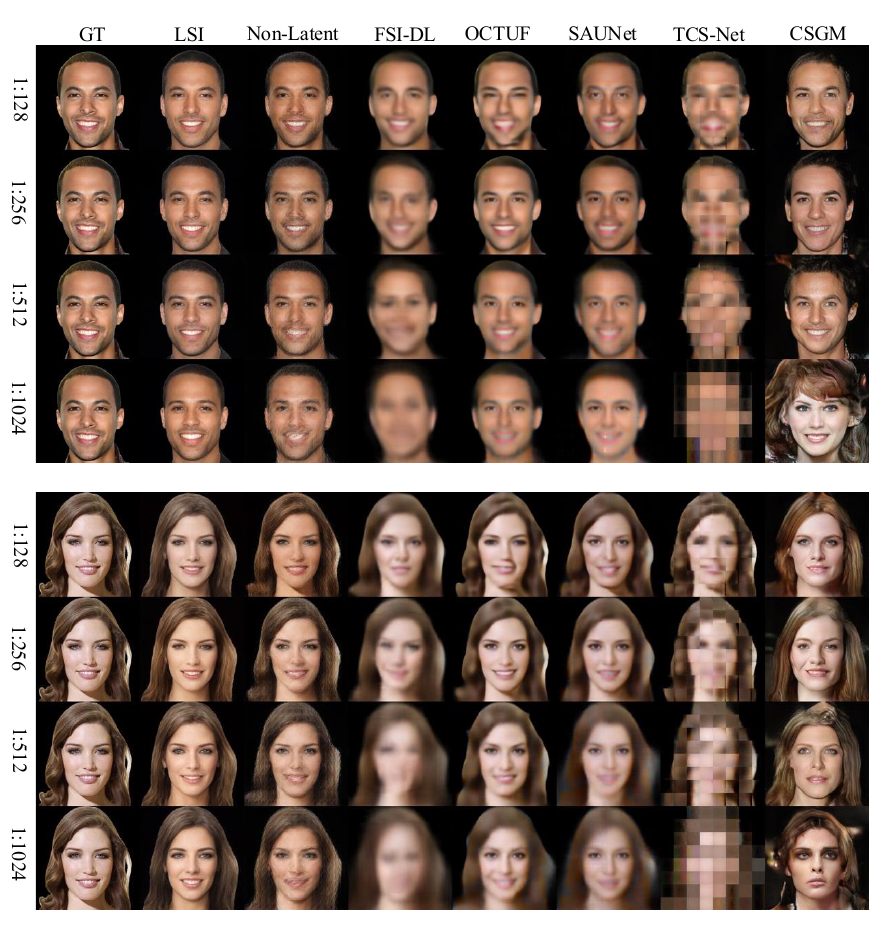}
    \caption{Different methods compared for facial reconstruction quality.}
    \label{fig:img3}
\end{figure*}

When evaluating under our heavily compressed setting, AuSamNet cannot optimize the mask, leading to unstable results. FSI-DL is capable of reconstructing the images but produces low-quality results as demonstrated by Figures~\ref{fig:img1}, \ref{fig:img2}, \ref{fig:img3}.

\paragraph{SAUNet~\cite{wang2023saunet}}

This approach proposes a 2D measurement system and an unfolding network as a reconstruction algorithm. Their measurement can be defined by the equation $Y = HXW^{T}$, where $H$ and $W$ are learned during the optimization process, and $X$ is the input. To enable a fair comparison, we clamped and then quantize their measurement matrix during training to be binary, following the same procedure adopted by our methods. We add a normalization layer after utilizing $H$ and $W$ to avoid exploding gradients and instability.

\paragraph{OCTUF~\cite{song2023optimization} and TCS-Net~\cite{tcsnet}} Both methods utilize transformer-based reconstruction techniques integrated with block-based image compressed sensing for measurements, where the image is divided into smaller patches. However, the sensing matrices used are not restricted to values below 1, making their physical implementation unfeasible. The methods were trained using the original code without any modifications, apart from adjustments to the compression ratios. Finally, they benefit from an unbounded sensing matrix and similar to SAUNet~\cite{wang2023saunet}, they do not model color, only learning the luminance.

\paragraph{Optimization-Based (CSGM~\cite{jalal2020robust})} leverages a generative model as a prior to reconstruct compressed signals. However, its measurement matrix is neither bounded nor quantized, making it unsuitable for physical deployment. Additionally, it requires multiple iterations and cannot achieve reconstruction with a single forward pass.

\paragraph{Non-Latent LSI} We further evaluate our approach by implementing a version of LSI that does not utilize the latent space. In this configuration, the 2D image $\in \mathbb{R}^{256 \times 256}$ is reconstructed directly from the measurements using Differential Ghost Imaging (DGI)~\cite{ferri2010differential}, followed by a state-of-the-art image reconstruction model~\cite{chu2022nafssr}, similar to the methodology described by~\cite{Wang:22}. This replaces the latent space representation and generative model entirely with a direct reconstruction pipeline. The training process, including perceptual and identity loss functions, Optical Encoder (\optenc) optimization, and other parameters, remain unchanged. While this version performs well for lower compression ratios (1:128), it is consistently outperformed by LSI, particularly at higher compression levels such as 1:512 and 1:1024 (see Table~\ref{tab:results_sup}), underscoring the critical role of latent space representation.

\begin{table}[ht]
    \centering
    \begin{tabular}{l r *{2}{>{\centering\arraybackslash}p{1.6cm}}}
        \toprule
        & VGGFace$\uparrow$ & Dlib$\uparrow$ & FID$\downarrow$ \\
        \midrule
        LSI Random  & 48.9\% & 71.7\% & 37.20 \\
        LSI & {\bf90.98}\% & {\bf92.68}\% & {\bf26.62} \\
        \bottomrule
    \end{tabular}
    \caption{Quantitative comparison between LSI and LSI with fixed random patterns (akin to classical Compressed Sensing), emphasizing the critical role of jointly optimizing the optical encoder for enhancing the performance. Compression ratio of 1:256.}
    \label{tab:random}
\end{table}

\subsection{Background Discussion}

To emphasize facial features, we mask out the background to prevent our already highly compressed signal from being used on non-facial information. It is worth noting, however, that our method remains effective even when trained on faces with uncontrolled backgrounds, achieving identity classification accuracies of $86.26\%$ and $86.88\%$ with VGGFace and DLib, respectively. Ablation study conducted for a compression ratio of 1:256.

\subsection{Optical Encoding vs. Fixed Random Encoding}

Optimizing the optical encoding plays a critical role in image reconstruction. This is evident in Table~\ref{tab:random}, which shows a significant performance drop when LSI is trained using fixed random patterns instead optimized encoding towards the latent space.


\end{document}